\documentclass{article}

\usepackage{arxiv}

\usepackage[utf8]{inputenc} 
\usepackage[T1]{fontenc}    
\usepackage{hyperref}       
\usepackage{url}            
\usepackage{booktabs}       
\usepackage{amsfonts}       
\usepackage{amsmath, amssymb}
\usepackage{nicefrac}       
\usepackage{microtype}      
\usepackage{graphicx}
\usepackage{float}
\usepackage{multirow}
\usepackage{array}
\usepackage{natbib}
\usepackage{doi}

\usepackage{algorithmicx}
\usepackage[ruled, vlined, linesnumbered]{algorithm2e}
\usepackage[noend]{algpseudocode}

\newcommand{\specialcell}[2][c]{%
\begin{tabular}[#1]{@{}c@{}}#2\end{tabular}}
\newcolumntype{P}[1]{>{\centering\arraybackslash}p{#1}}
\newcolumntype{M}[1]{>{\centering\arraybackslash}m{#1}}

\SetKwBlock{Begin}{Function}{end}
\DontPrintSemicolon 

\DeclareUnicodeCharacter{2212}{-}
\DeclareUnicodeCharacter{02BC}{-}

\title{Using deep generative neural networks to account for model errors in Markov chain Monte Carlo inversion}

\author{ \href{https://orcid.org/0000-0002-6981-3232 }{\includegraphics[scale=0.06]{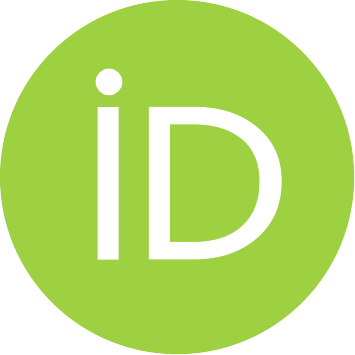}\hspace{1mm}Shiran Levy}\\
    Institute of Earth Sciences \\
	University of Lausanne \\
	Lausanne, Switzerland \\
	\And
	Jürg Hunziker\\
    Institute of Earth Sciences \\
	University of Lausanne \\
	Lausanne, Switzerland \\
	\And	
	Eric Laloy\\
    Belgian Nuclear Research Center \\
	Mol, Belgium \\
	\And
	James Irving\\
    Institute of Earth Sciences \\
	University of Lausanne \\
	Lausanne, Switzerland \\
	\And
	Niklas Linde\\
    Institute of Earth Sciences \\
	University of Lausanne \\
	Lausanne, Switzerland \\
}

\date{\small{Original publication: Geophys. J. Int., 23 September 2021\\
DOI: \url{https://doi.org/10.1093/gji/ggab391}}}

\renewcommand{\shorttitle}{Using deep generative neural networks to account for model errors}

\hypersetup{
pdftitle={Using deep generative neural networks to account for model errors in Markov chain Monte Carlo inversion},
pdfsubject={physics.geo-ph},
pdfauthor={Shiran Levy, Jürg Hunziker, Eric Laloy, James Irving, Niklas Linde},
pdfkeywords={Inverse theory, Neural networks, Ground penetrating radar, Probability distributions, Hydrogeophysics},
}

\begin{document}
\maketitle

\begin{abstract}
Most geophysical inverse problems are nonlinear and rely upon numerical forward solvers involving discretization and simplified representations of the underlying physics. As a result, forward modeling errors are inevitable. In practice, such model errors tend to be either completely ignored, which leads to biased and over-confident inversion results, or only partly taken into account using restrictive Gaussian assumptions. Here, we rely on deep generative neural networks to learn problem-specific low-dimensional probabilistic representations of the discrepancy between high-fidelity and low-fidelity forward solvers. These representations are then used to probabilistically invert for the model error jointly with the target geophysical property field, using the computationally-cheap, low-fidelity forward solver. To this end, we combine a Markov-chain-Monte-Carlo (MCMC) inversion algorithm with a trained convolutional neural network of the spatial generative adversarial network (SGAN) type, whereby at each MCMC step, the simulated low-fidelity forward response is corrected using a proposed model-error realization. Considering the crosshole ground-penetrating radar traveltime tomography inverse problem, we train SGAN networks on traveltime discrepancy images between: (1) curved-ray (high fidelity) and straight-ray (low fidelity) forward solvers; and (2) finite-difference-time-domain (high fidelity) and straight-ray (low fidelity) forward solvers. We demonstrate that the SGAN is able to learn the spatial statistics of the model error and that suitable representations of both the subsurface model and model error can be recovered by MCMC. In comparison with inversion results obtained when model errors are either ignored or approximated by a Gaussian distribution, we find that our method has lower posterior parameter bias and better explains the observed traveltime data. Our method is most advantageous when high-fidelity forward solvers involve heavy computational costs and the Gaussian assumption of model errors is inappropriate. Unstable MCMC convergence due to nonlinearities introduced by our method remain a challenge to be addressed in future work.  
\end{abstract}


\section{Introduction}\label{sec:intro}

Bayesian inversion treats model parameters as random variables that are constrained by prior probability density functions and noise-contaminated data through a likelihood function \citep{tarantola2005inverse,gelman2013bayesian}. The Bayesian framework is flexible in that it allows accounting for uncertainties due to inaccurate or incomplete descriptions of the underlying physics of the problem, as well as for errors related to the measurement process. We refer to the former as model errors \citep{kaipio2007statistical} because they describe inaccuracies in the forward modeling used to connect physical properties to observable data, while other authors have used the term "theoretical error" \citep{tarantola1982inverse} in a similar context. Model errors are notoriously difficult to quantify, particularly when the forward problem at hand is nonlinear. Their magnitudes and correlation patterns can be highly complex and variable throughout the model parameter space, and deriving an appropriate statistical description of them is therefore challenging. At the same time, relying on accurate state-of-the-art forward solvers with minimal model errors is not always practical as they are generally computationally expensive, which becomes particularly problematic when the forward response has to be calculated many times. Surrogate models (also referred to as proxy models or low-fidelity models) implying an approximation or a simplified representation of the underlying physical process offer an attractive alternative provided that one can adequately account for the resulting model errors. Model errors are commonly an order of magnitude or so larger than measurement uncertainties \citep{tarantola1982inverse,kaipio2007statistical, hansen2014accounting}. Therefore, ignoring them might lead to severe bias, artifacts and over-confident results \citep{brynjarsdottir2014learning, hansen2014accounting}.

Early pioneering work on model errors was conducted by \citet{kennedy2001bayesian}. They represent model errors as a Gaussian process (GP) that is conditioned at locations in the model parameter space where the model errors are known. The general applicability of this method for geoscientific inverse problems of high dimensional and multivariate nature remains unclear \citep{linde2017uncertainty} even if some promising applications exist \citep{xu2015bayesian,xu2017quantifying}. Most approaches dealing with model errors involve building a statistical model of the discrepancy between a high-fidelity forward model and a cheaper, less-accurate counterpart. Some of these methods formulate the likelihood function such that prior knowledge about the mean and covariance of the model errors is incorporated \citep{kaipio2007statistical, cui2011bayesian, hansen2014accounting}. Despite their proven value, the Gaussian assumptions made in these methods might be problematic when confronted with non-Gaussian priors, non-Gaussian observational noise and nonlinear problems. Traditionally, model errors are learned by evaluating modeling discrepancies using samples from the prior, yet, recent adaptive approaches in which the model error description is updated based on samples from the posterior region has shown important improvements \citep{cui2011bayesian, calvetti2014dynamic}. Other approaches for dealing with model errors involve estimating and removing them from the residual data term before calculating the likelihood function \citep{kopke2018accounting, kopke2019stochastic}. In such methods, the residuals are projected onto an orthogonal model-error basis, which is constructed either during the inversion using a dictionary-based K-nearest-neighbour approach, or before the inversion using principal component analysis (PCA) conducted on a suite of model-error realizations. The dynamic model error estimation methods of \citet{cui2011bayesian}, \citet{calvetti2014dynamic} and \citet{kopke2018accounting} enjoy local statistics of model errors in regions of high posterior density; however, they do require occasional runs of a high-fidelity forward solver during the inversion. Another approach is presented by \citet{rammay2019quantification} who perform joint inversion of the model parameters and error-model in the context of reservoir history matching. They use PCA basis functions to parameterize the error-model and infer for the PCA coefficients during inversion. 

Over the past decade, the use of machine learning (ML) in geophysical applications has become increasingly popular as a result of continuing growth in computational resources and numerous breakthroughs in ML research \citep{giannakis2019machine, bergen2019machine, dramsch202070, yu2020data}. Deep learning models, an extension to machine learning models, can be trained to produce an amortized data-based alternative to expensive physics-based models \citep{TRIPATHY2018565,TANG2020109456,JIN2020107273}. Nonetheless, these models are problem specific and their accuracy may vary depending on availability of training data and their ability to generalize. Furthermore, as surrogate models they still suffer from some degree of model errors when compared to the high-fidelity model which they aim to approximate. Here we give several examples of machine learning applications addressing model errors. The approach of \citet{xiao2019error}, in analogy to the GP approach of \citet{kennedy2001bayesian}, uses GP regression, an ML algorithm, to learn a set of error response functions associated with a low-fidelity flow model. The error response functions predict a set of parameters that through proper orthogonal decomposition are projected into the full error space and used to correct the low-fidelity model. \citet{seille2020bayesian} utilize regression trees in order to learn a dimensionality discrepancy model (DDM) predicting the model errors associated with using 1D instead of 3D magnetotelluric modeling. The DDM is then used to define a likelihood function that is used within a reversible-jump Markov chain Monte Carlo (MCMC) procedure \citep{green1995reversible}. \citet{sun2019combining} apply convolutional neural networks (CNN's) describing spatial and temporal discrepancies between land surface model (LSM) predictions and observations from the gravity recovery and climate experiment (GRACE). Their neural network combining three CNN architectures receives as an input the LSM output as well as additional predictors (precipitation and temperature) and in return outputs the mismatch between the LSM and GRACE observations. Their study shows an increased correlation between corrected LSM and observed data, thereby, highlighting the potential of deep-learning to improve geo-scientific models over different spatiotemporal scales. Machine and deep-learning algorithms have also been proven efficient for parameterizing geological models \citep{laloy2017inversion,laloy2018training,mosser2020stochastic}. \cite{laloy2018training} parameterized model realizations using a spatial generative adversarial network (SGAN) and integrated the generating part of the network within an MCMC routine. In this type of neural network, a nonlinear transformation is learned using training images. The image space, representing the high-dimensional space on which forward simulations are performed, is connected to a lower dimensional space (latent space) through a series of nonlinear transformations in the form of convolution operations. The inversion is performed with respect to this lower-dimensional representation. Given the notable reduction in the number of inferred parameters, the spatial nature of the network and the fast generation of model realizations, the SGAN-parameterization was able to significantly improve the MCMC inversion performance compared with sequential geostatistical resampling \citep{mariethoz2010bayesian,ruggeri2015systematic}.

In this study, we use SGANs to learn a low-dimensional parameterization of model errors associated with a low-fidelity forward solver. A notable characteristic of the SGAN is its localized nature, allowing for perturbations in a specific region of the image space following a perturbation in one of the latent parameters. Our approach takes advantage of spatial correlation within model-error realizations to transform the high-dimensional model-error space (same dimension as the data space) into a lower-dimensional latent space. We train two separate deep generative neural networks, one for the subsurface model parameters and the other for the model errors. Then, we perform MCMC inversion on the latent parameters to infer the joint posterior distribution of both. We consider numerical simulations in the context of crosshole ground-penetrating radar (GPR) traveltime tomography and test our method with synthetic data generated by either a (1) curved-ray (eikonal) or (2) finite difference time-domain full-waveform forward solver. The inversion on the other hand is performed using a low-fidelity straight-ray forward solver. The aim of our approach is to account for discrepancies in the modelling process when one replaces an expensive, high-fidelity solver with a cheap, less accurate solver to speed up the inversion process. By doing so, we hope to reduce the bias caused by using low-fidelity solvers while allowing for an efficient MCMC inversion. Note that the cheap low-fidelity solver could, in principle, also be a deep-learning based forward solver that was trained on the same database of high-fidelity forward solvers. We compare our approach against two alternative inversion approaches that also rely on the same low-fidelity forward solver, one where model errors are ignored and the other where they are approximated as Gaussian. For the case of the synthetic data being generated with the eikonal solver, we also compare with inversion results obtained without any model errors, that is, when using the eikonal solver as forward model in the MCMC inversions.


\section{Methods}

Our approach to account for model errors involves three main steps: (1) database preparation, (2) SGAN training, and (3) MCMC inversion. The database preparation involves setting up the database on which the neural networks for the subsurface model parameters and model error are trained. During training, information about the trained parameters of the generative network is given at regular intervals. The stage (generator iteration) at which training data are retained to generate realizations for subsequent inversions is chosen according to statistical measures as well as visual inspection. Finally, the deep generative neural networks are integrated into an MCMC inversion algorithm. Below we describe each of the three stages in detail in the context of the considered crosshole GPR traveltime tomography inverse problem.

\subsection{Database preparation}
\subsubsection{Multi-Gaussian model database}\label{sec:model}

The training image (TI) used as a basis to describe the spatial structure of our subsurface model-parameter prior is a $2500 \times 2500$ pixels, ($250 \times 250$ m) anisotropic, multi-Gaussian, geostatistical realization with a variance of 1 and mean of zero. It was generated by \citet{pirot2017probabilistic} based on the  geostatistical analysis of sediments at the Boise Hydrogeophysical Research Site conducted by \citet{barrash2002hierarchical}. We split the TI into two parts: a segment of size $2250 \times 2500$ pixels, which is used for training the SGAN, and a segment of size $250 \times 2500$ pixels, from which we select the reference models used in our inversion examples. The training is performed on small patches $\mathbf{X}_{\boldsymbol{\Phi}}$ of pre-defined size which are randomly cropped from the segment of the TI intended for training. The porosity field $\boldsymbol{\Phi}$ is then computed from the multi-Gaussian realizations using the lognormal transformation 

\begin{equation}\label{eq:por}
\boldsymbol{\Phi} = \exp(\mathbf{X}_{\boldsymbol{\Phi}}\times 0.22361 - 1.579)
\end{equation}

in \citet{pirot2017probabilistic}.

\subsubsection{Crosshole GPR simulations and model-error database}\label{subsec:moderr}

In a crosshole GPR experiment, an electromagnetic impulse is emitted from a source antenna located in one borehole and registered in a receiver antenna positioned in an adjacent borehole. To create a model-error database of first-arrival travel time residuals, we perform crosshole GPR numerical simulations based on the ${\boldsymbol{\Phi}}$-realizations  described in subsection \ref{sec:model} using the low- and high-fidelity forward solvers, which we denote by $g^{LF}$ and $g^{HF}$, respectively. The numerical simulations are performed on slowness $\textbf{s}$ (1/velocity) fields, therefore, the porosity field of the subsurface model-parameter realizations must be converted to a slowness field. This can be done via the following relationships \citep{pride1994governing}:

\begin{equation}\label{eq:keff}
    \boldsymbol{\kappa}_{b} = \boldsymbol{\Phi}^{m}\kappa_{w}+(1-\boldsymbol{\Phi}^{m})\kappa_{s},
\end{equation}

and 

\begin{equation}\label{eq:slow}
    \mathbf{s} = \frac{\sqrt{\boldsymbol{\kappa}_{b}}}{c}, 
\end{equation}

\noindent
where $\kappa_{w}$ and $\kappa_{s}$ are the water and rock dielectric constants, $m$ is the cementation exponent, $\boldsymbol{\kappa}_{b}$ ($b$ stands for "bulk") is the effective dielectric constant of the medium and $c$ is the speed of light in vacuum. We ignore petrophysical prediction uncertainty related to scatter in the petrophysical relationship \citep{brunetti2018impact} and assume the petrophysical parameters to be known. Following \citet{pirot2017probabilistic}, we set $\kappa_{w}$ to be $81$, $\kappa_{s}$ to $6$ and $m$ to $1.48$. 

Assuming that the forward solver $g^{HF}$ (HF stands for high-fidelity) describes perfectly the crosshole GPR experiment, we have:

\begin{equation}\label{eq:fw_hf}
    \textbf{d} = g^{HF}(\textbf{s}) + \boldsymbol{\epsilon},
\end{equation}

\noindent
where $\textbf{d}$ represents the observed traveltime data corresponding to slowness parameters $\textbf{s}$ with observational noise $\boldsymbol{\epsilon}$. The proxy solver $g^{LF}$ gives rise to a model error $\eta(\textbf{s})$:

\begin{equation}\label{eq:fw_lf}
    \textbf{d} = g^{LF}(\textbf{s}) + \eta(\textbf{s}) + \boldsymbol{\epsilon},
\end{equation}

\noindent
describing the discrepancy between the two solvers for each source-receiver pair:

\begin{equation}\label{eq:err} 
    \eta(\textbf{s}) = g^{HF}(\textbf{s}) - g^{LF}(\textbf{s}).
\end{equation}

To test our method, we consider two different model errors for the crosshole GPR traveltime tomography problem. In both test cases, we use a straight-ray solver denoted by $g^{SR}$ as our low-fidelity solver $g^{LF}$. In the first test case, we consider a finite difference approximation of the eikonal equation by \citet{podvin1991finite} as the high-fidelity model, such that $g^{HF} = g^{eikonal}$ and the model error is $\eta^{eikonal-SR}(\textbf{s}) = g^{eikonal}(\textbf{s}) - g^{SR}(\textbf{s})$. In the second test case, the high-fidelity model is based on a finite difference time-domain scheme (FDTD) (\citealp{irving2006numerical}), such that $g^{HF} = g^{FDTD}$ and the model error is $\eta^{FDTD-SR}(\textbf{s}) = g^{FDTD}(\textbf{s}) - g^{SR}(\textbf{s})$. We note that our method is almost fully amortized as the computationally expensive high-fidelity solver is only used prior to inversion to create the model-error database and, in a synthetic example such as ours, the data (observed data) that are to be inverted. 

From the FDTD simulations, the first-arrival travel times are automatically chosen by identifying the first maximum of the signal and subtracting the time delay between the source wavelet's initiation and first peak. Due to an underlying infinite-frequency assumption, ray-based approaches (straight ray and eikonal solvers) provide the same arrival times in 2D and 2.5D media. This is not the case for FDTD simulations leading to important time shifts in the 2D FDTD first-break picks compared to the ray-based solvers. We correct this phase shift by applying a reversed geometrical correction to that found in \citet{ernst2007application}, effectively performing a 2D to 2.5D correction of the FDTD data:

\begin{equation}\label{eq:geoCorrection}
\hat{E}(\textbf{x}_{trn},\textbf{x}_{rec},\boldsymbol{\omega}) =  \frac{E(\textbf{x}_{trn},\textbf{x}_{rec},\boldsymbol{\omega})}{\sqrt{\frac{2\pi T(\textbf{x}_{trn},\textbf{x}_{rec})}{-i\boldsymbol{\omega} \bar{\kappa} \mu_{0}}}},
\end{equation}

\noindent
where $E(\textbf{x}_{trn},\textbf{x}_{rec},\boldsymbol{\omega})$ and $\hat{E}(\textbf{x}_{trn},\textbf{x}_{rec},\boldsymbol{\omega})$ are the signal in the frequency domain before and after applying the correction from 2D to 2.5D, respectively, for source and receiver locations $\textbf{x}_{trn}$ and $\textbf{x}_{rec}$. Here, $T(\textbf{x}_{trn},\textbf{x}_{rec})$ are the picked arrival times based on signal $E(\textbf{x}_{trn},\textbf{x}_{rec})$ in the time domain, $\boldsymbol{\omega}$ refers to the angular frequency of the signal, $\bar{\kappa}$ is the mean dielectric constant of the medium, $\mu_{0}$ is the magnetic permeability in vacuum and $i^{2}=−1$. After correction, arrival times were repicked on the corrected signals. 

\subsection{Generative adversarial networks} \label{sec:SGAN}

In a fully connected neural network (see \cite{goodfellow2016deep} for details), a single neuron with weight vector $\mathbf{w}$, bias term $b$, and input vector $\mathbf{x}$ can be represented as

\begin{equation}\label{eq:nn}
h(\mathbf{x};\mathbf{w},b) = \varphi (\sum^{N_{x}}_{i=1} w_{i}x_{i} +b),
\end{equation}

\noindent
where $\varphi$ is a nonlinear transformation referred to as the activation function. In a convolutional neural network applied to an image, a single pixel at location $(u,v)$ in the output feature map $\textbf{F}$ is a result of a convolution between a kernel $\textbf{K}$ of size $N_{H} \times N_{W}$ and a sub-region of the same size in the input image $\textbf{I}$:

\begin{equation}\label{eq:cnn} 
        \textbf{F}_{u,v} = \varphi( \sum^{N_{W}}_{j = 1} \sum^{N_{H}}_{i = 1} \textbf{K}_{i,j}\textbf{I}_{u+i,v+j} + b).
\end{equation}

\noindent
The full feature map is the collection of pixels resulting from convolution operations over different locations in the input image. A convolutional layer produces multiple feature maps, each being a result of convolution between the input image and a different filter. All filters in a layer share the same dimensions, but contain different weights. A deep convolutional network is a network in which several convolutional layers are sequentially stacked. As the number of layers and neurons within layers increases, the ability of the network to express complex functions increases.

\begin{figure}
    \centering
    \includegraphics[trim = 0 0 0 0,  clip,width=0.82\textwidth]{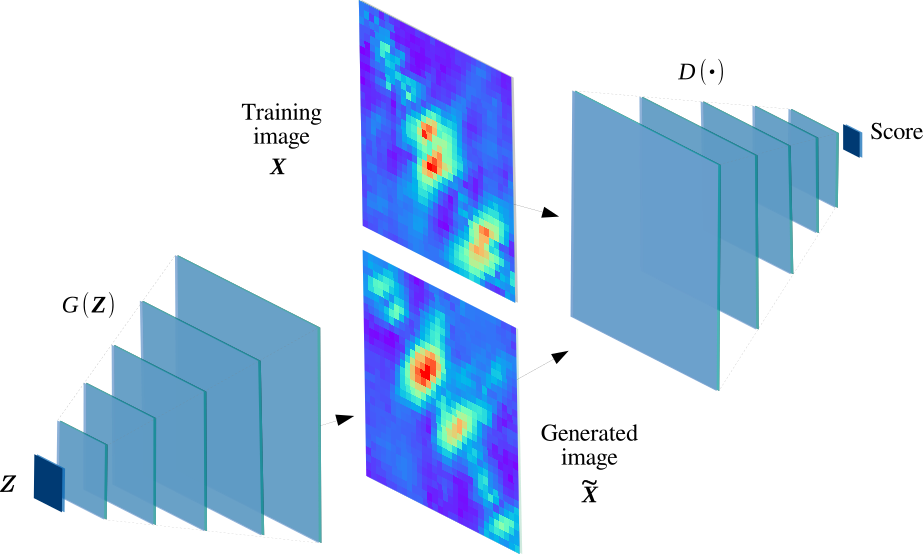}
    \caption{Illustration of our SGAN architecture with five layers when applied to represent model errors. During training, each parameter in the latent space $\textbf{Z}$ is randomly drawn from a uniform distribution $\mathcal{U}(-1,1)$. Each $\textbf{Z}$ is transformed into a single image $\tilde{X}$ through a nonlinear transformation $G(\cdot)$. At each iteration, a batch of images $\tilde{X}$ (generated) and a batch of images $X$ (training) are interchangeably fed into the critic $D(\cdot)$, resulting in a score that is then used to update the network parameters through back-propagation.}
    \label{fig:SGAN_arch}
\end{figure}

A generative adversarial network (GAN; \citealp{goodfellow2014generative}) is a convolutional neural network (CNN), in which training is a zero-sum game between a generator $G$ and a discriminator $D$. The GAN seeks to minimize a distance between the distribution $P_{r}$ of the training data and the distribution $P_{g}$ of the data created by the generator $G$. The generator input is usually a low-dimensional latent vector $\textbf{z}$ drawn from a uniform distribution $\mathcal{U}(-1,1)$ or a standard normal distribution $\mathcal{N}(0,1)$, and the output is an image $\mathbf{\tilde X}$. \citet{jetchev2016texture} extended the GAN into a spatial-GAN (SGAN), where the input $\textbf{Z}$ becomes a 2D (later extended to 3D by \citealp{laloy2018training}) tensor of $n \times m$ ($\times q$) dimensions such that a perturbation in one tensor element corresponds to a change in a specific region of the output image $\mathbf{\tilde X}$. The input to the discriminator $D$ is either an image $\mathbf{\tilde X}$ from the generator distribution $P_{g}$ or an image $\textbf{X}$ from the training distribution $P_{r}$ (see Fig. \ref{fig:SGAN_arch}). At each training iteration, a batch of generated images $\mathbf{\tilde X}$, and a batch of training images $\textbf{X}$ are interchangeably fed into the discriminator and, according to the loss function in use, they are either classified as $0$ (fake) or $1$ (true), or are given a score. As opposed to other types of deep generative networks (e.g. variational autoencoders), training enforces only the distribution on $\mathbf{\tilde X}$ ($P_{g}$) to approximate the distribution on $\mathbf{X}$ ($P_{r}$) while the prior on $\textbf{Z}$ is simply assigned such that, for example, all draws during training are drawn from a uniform distribution $\mathcal{U}(-1,1)$. For an enhanced stability of training and better general performance, we use the Wasserstein loss function \citep{arjovsky2017wasserstein}, whereby the distance between distributions $P_{r}$ and $P_{g}$ is based on the Wasserstein-1 distance $W(P_{r},P_{g})$: 

\begin{equation}\label{lossf}
    \min_{G} \max_{D \in \mathcal{D}} \underset{\textbf{X} \sim P_{r}}{\mathbb{E}} [D(\textbf{X})] - \underset{\textbf{Z} \sim p_{g}}{\mathbb{E}} [D(G(\textbf{Z}))].
\end{equation} 

\noindent
Given that the output of $D(\cdot)$ in equation \eqref{lossf} is a score rather than a classification to $0$ and $1$, it is referred to as a "critic". Once gradients of the loss function are calculated with respect to the network parameters, the error is back-propagated through the network, allowing updates of the weights and biases of each layer.

\subsection{SGAN architecture and training}\label{sec:training}

The network architecture of the generator and critic are asymmetric with respect to each other (see Appendix \ref{ap:lr} for details) and each of them contains five sequentially stacked convolutional layers. Spectral normalization is applied to the weights in each critic layer \citep{miyato2018spectral}. This normalizes the weight matrices $\textbf{K}$ with respect to the spectral norm at each layer, thus forcing them to conform to the Lipschitz continuity condition. In the SGAN trained on model errors, we apply mean spectral normalization to critic layers (\citealp{subramanian2019mean}) as it improved the quality of the generated model-error realizations. The generator feature maps are normalized with respect to features (elements) using instance normalization (\cite{ulyanov2016instance}). The first four layers of the critic and the generator are followed by a rectified linear unit (ReLU):$f(x) = max(0,x)$ and a LeakyReLU: $f(x) = max(0.2x,x)$ activation function, respectively, and the last layer in the generator is followed by a tanh activation function. We set the learning rates of the generator and critic according to the two time-scale update rule (TTUR) by \citet{heusel2017gans}, with a ratio of $1:4$. The output size $x$ of layers $l=1,..,5$ in the generator can be calculated via the following relationship:

\begin{equation}\label{eq:output}
x_{l} = s \cdot (x_{l-1}-1)-2 \cdot p + (k - 1) + 1,    
\end{equation}

\noindent
where $s$ is the stride controlling movement of the filter along the image, $p$ is the the number of padding columns/rows of zeros added to the layer's input, and $k$ is the kernel size (see \cite{dumoulin2016guide} for more information). We use padding to control the output size and obtain an image with dimensions that are as close as possible to our model size (see Appendix \ref{Ap:training} for more details).

All images fed into the critic must be normalized to a $\left[-1,1\right]$ range and have the same dimensions. Thus, TI's are either cropped (subsurface model parameter) or linearly interpolated (model error) into a size fitting that of the generative network's output. After training, the generated subsurface model parameter $\mathbf{\tilde X}_{\boldsymbol{\Phi}}$ or model-error $\mathbf{\tilde X}_{\boldsymbol{\eta}}$ realizations are either cropped or interpolated to the desired image size and re-scaled back to the original value range. In the case of the subsurface model-parameter realizations there is an additional step where porosity values are assigned according to equation \ref{eq:por}.

\subsection{Bayesian inference of latent parameters}\label{bayes_inv}

We aim to estimate the low-dimensional (latent-space) representation of the subsurface model parameters and associated model error by incorporating the two trained generative networks within an MCMC inversion. Subsurface model-parameter and model-error prior realizations are generated using the SGAN and the forward responses during inversion are computed using the straight-ray solver $g^{LF} = g^{SR}$. The posterior probability density function (pdf) $p(\textbf{Z}|\textbf{d})$ is expressed through Bayes' theorem as:

\begin{equation}
    p(\textbf{Z}|\textbf{d}) = \frac{p(\textbf{d}|\textbf{Z})p(\textbf{Z})}{p(\textbf{d})},
\end{equation}

\noindent
where $p(\textbf{d}|\textbf{Z})$ is the likelihood function, $p(\textbf{Z})$ is the prior pdf of latent parameters $\textbf{Z}$, and $p(\textbf{d})$ is the marginal likelihood (evidence). The latter is a constant that we ignore in this work and we thus focus on the unnormalized posterior $p(\textbf{Z}|\textbf{d}) \propto  p(\textbf{d}|\textbf{Z})p(\textbf{Z})$. For numerical reasons we work with the log-likelihood which, assuming the measurement errors are independent, identical and normally distributed, is given by:

\begin{equation}\label{eq:logL}
        l(\textbf{d}|\textbf{Z}) = -\frac{N_{d}}{2}\log(2\pi)-N_{d}\log(\sigma)-\frac{1}{2}\sigma^{-2} \left[\textbf{d}-\textbf{d}_{sim} \right]^2,
\end{equation}

\noindent
where $N_{d}$ is the number of data points, $\sigma$ is the standard deviation of the measurement errors $\boldsymbol{\epsilon}$, and $\textbf{d}_{sim}$ and $\textbf{d}$ are the forward simulated and observed data, respectively. To sample from the posterior distribution, we rely on the differential evolution adaptive Metropolis (DREAM$_{(ZS)}$) algorithm, in which MCMC chains evolve in parallel and jumps are proposed based on candidate points from an archive of past states \citep{ter2008dreamZS, vrugt2009dreamZS, laloy2012high}. In this algorithm, the jump size is given by $\gamma = \frac{2.38}{\sqrt{2\delta d^{'}}} \beta $, where $\beta$ is a user defined scalar referred to here as the jump rate scaling factor, $\delta$ is the number of candidate points pairs used to generate the proposal, and $d^{'}$, the number of dimensions to be updated, varies during the inversion according to a crossover (CR) probability \citep{laloy2012high}. At each MCMC step and for each individual chain, a random sample is drawn from the proposal distribution $q(\textbf{Z}\sp{\prime},\textbf{Z}^{t-1})$, which is symmetric with boundary handling to ensure that the samples are drawn proportionally to the uniform prior. As the prior is uniform, the sample is either accepted or rejected according to a transition acceptance rule $p_{acc}(\textbf{Z}^{t-1} \xrightarrow{} \textbf{Z}\sp{\prime}) = e^{(l(\textbf{d}|\textbf{Z}\sp{\prime}) - l(\textbf{d}|\textbf{Z}^{t-1}))}$. If accepted, the chain moves to $\textbf{Z}\sp{\prime}$ such that $\textbf{Z}^{t} = \textbf{Z}\sp{\prime}$. If rejected, the chain remains at the current sample and $\textbf{Z}^{t} = \textbf{Z}^{t-1}$. We run the inversion with eight parallel chains and, to improve the search, we allow for a $20\%$ chance of snooker update \citep{ter2008dreamZS} during the first $20,000$ steps (per chain) which we consider as the burn-in period. As opposed to parallel updating where sampling occurs along an axis that runs past states of a single chain, the snooker update involves an axis that runs along states of two different chains. The jump rate scaling factor $\beta$ is varied adaptively during the burn-in period in order to reach a $20\% - 30\%$ MCMC acceptance rate. To prevent very high acceptance rates and slow mixing after the burn-in period, we set a minimum value to the $\beta$, beyond which it cannot decrease.

\begin{figure}
    \centering
    \includegraphics[trim = 50 40 50 20,  clip,width=0.85\textwidth]{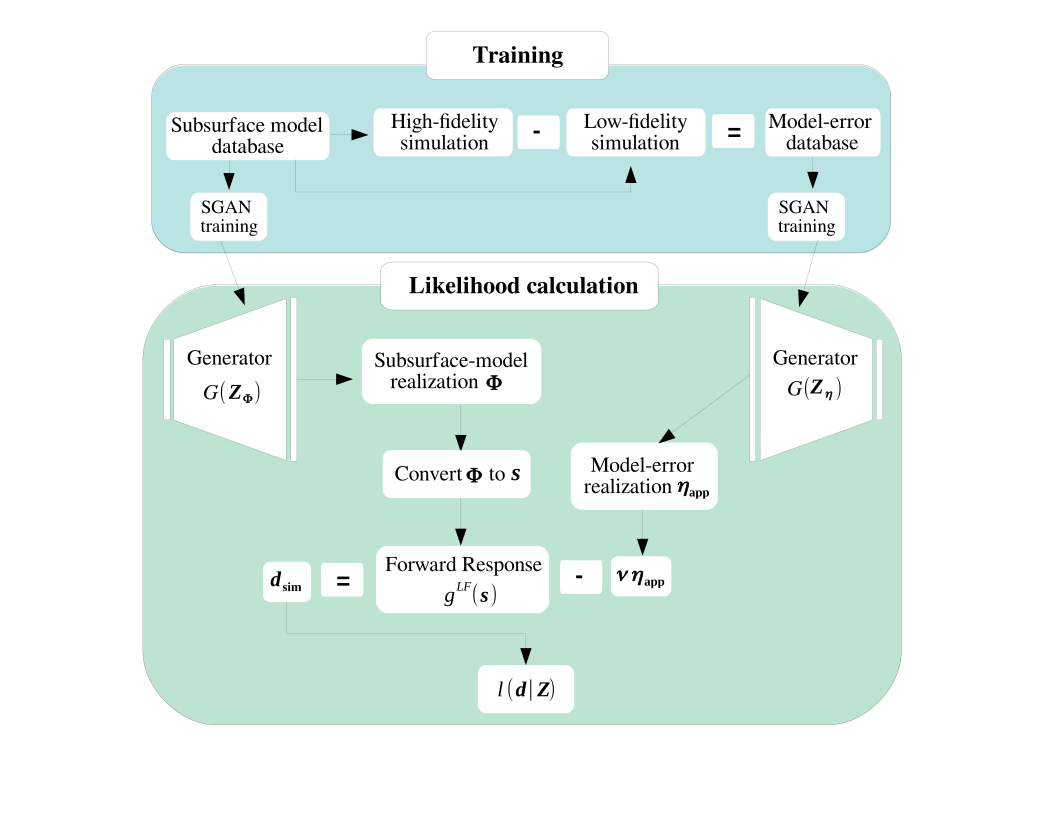}
    \caption{SGAN-ME workflow. Subsurface-model representation using SGAN is discussed in details in the work of \citet{laloy2018training}, here we focus on model-error representation.}
    \label{fig:SGAN_workflow}
\end{figure}

We jointly infer the posterior distribution of the two low-dimensional latent spaces: $\textbf{Z}_{\boldsymbol{\Phi}}$ describing the subsurface model parameters and $\textbf{Z}_{\boldsymbol{\eta}}$ describing the model error, both of which have uniform prior distributions $\mathcal{U}(-1,1)$. The proposed latent parameter realizations are mapped into their respective high-dimensional image spaces $\boldsymbol{\Phi}$ and $\boldsymbol{\eta}_{app}$ (approximate model error), where a low-fidelity forward response is calculated on the porosity field $\boldsymbol{\Phi}$ converted to slowness $\textbf{s}$ using equations \eqref{eq:keff} and \eqref{eq:slow}. In addition to the subsurface model-parameter and model-error latent parameters, we infer an auxiliary parameter $\nu$ with a uniform prior distribution $\mathcal{U}(0,1)$ that scales the model-error realization before it is added to the simulated data. This scalar was found to improve the inference and quality of the inferred model errors by providing additional means to control their magnitudes. When inferring model errors, $\textbf{d}_{sim}$ in equation \eqref{eq:logL} is replaced with $g^{SR}(\textbf{s}) + \nu \boldsymbol{\eta}_{app}$. The most salient features of our method, combining SGAN-ME (ME stands for model error) with MCMC inversion, is provided in Algorithm \ref{Algorithm:MCMC_SGAN} and Figure \ref{fig:SGAN_workflow}.

We compare SGAN-ME against cases where model errors are zero as the high-fidelity forward solver is used in MCMC inversions or model errors are either ignored or approximated to be Gaussian. In these latter cases, the inferred parameters are the latent parameters of the model alone, such that $\textbf{Z} = \textbf{Z}_{\boldsymbol{\Phi}}$ and we simply plug $\textbf{d}_{sim} = g^{SR}(\textbf{s})$ into equation \eqref{eq:logL}.

To approximate the model errors as Gaussian, we follow \citet{hansen2014accounting} and learn their mean $\textbf{d}_{ME}$ and a covariance matrix $\textbf{C}_{ME}$, which are used to correct the residual term and inflate the likelihood function:

\begin{equation}\label{eq:hansen}
        l(\textbf{d}|\textbf{Z})  = - \frac{N_{d}}{2}  \log(2\pi)-\frac{1}{2}\log(|\textbf{C}_{D}|) -\frac{1}{2}  \left[\textbf{d}-g^{SR}(\textbf{s}) - \textbf{d}_{ME} \right]^{T}\textbf{C}_{D}^{-1}\left[\textbf{d}-g^{SR}(\textbf{s}) - \textbf{d}_{ME} \right],    
\end{equation}

\noindent
where $\textbf{C}_{D} = \textbf{C}_{d} + \textbf{C}_{ME}$, with $\textbf{C}_{d}$ being the traditional data covariance matrix and $\textbf{C}_{ME}$ the learned model-error covariance matrix. The bias correction term $\textbf{d}_{ME}$ is the model-error mean. We use $800$ random model-error samples from the same database used for training the SGAN to learn $\textbf{C}_{ME}$ and $\textbf{d}_{ME}$, noting that \cite{hansen2014accounting} recommend to use at least $300$ samples.

\vspace{0.2cm}%
\begin{algorithm}[H]
    \caption{SGAN-ME inversion with differential evolution adaptive Metropolis DREAM$_{(ZS)}$}\label{Algorithm:MCMC_SGAN}
    \small
        \SetAlgoLined
            Set \textit{t} = 1 and initialize the archive with realizations $\textbf{Z}_{\boldsymbol{\Phi}}$, $\textbf{Z}_{\boldsymbol{\eta}}$ and $\nu$ randomly drawn from $p(\textbf{Z}_{\boldsymbol{\Phi}})$, $p(\textbf{Z}_{\boldsymbol{\eta}})$ and $p(\nu)$ (respectively)\\
        Initialize $\textbf{Z}^{t} = [\textbf{Z}^{t}_{\boldsymbol{\Phi}} , \textbf{Z}^{t}_{\boldsymbol{\eta}},\nu^{t}]$ for each MCMC chain\\
        $\mathbf{\tilde X}_{\boldsymbol{\Phi}}^{t}$, $\mathbf{\tilde X}_{\boldsymbol{\eta}_{app}}^{t} \gets$  $\text{G}_{\boldsymbol{\Phi}}(\textbf{Z}_{\boldsymbol{\Phi}}^{t})$, $\text{G}_{\boldsymbol{\eta}}(\textbf{Z}_{\boldsymbol{\eta}}^{t})$\\
        Perform post-processing (section \ref{sec:training}):
        $\boldsymbol{\Phi}^{t}$, $\boldsymbol{\eta}_{app}^{t} \gets \mathbf{\tilde X}_{\boldsymbol{\Phi}}^{t}$, $\mathbf{\tilde X}_{\boldsymbol{\eta}_{app}}^{t}$ and
        convert $\boldsymbol{\Phi}^{t}$ into slowness $\textbf{s}^{t}$ (equations \eqref{eq:keff}-\eqref{eq:slow})\\
        $\mathbf{d}_{sim} = g^{LF}(\mathbf{s}^{t}) + \nu^{t} \boldsymbol{\eta}_{app}^{t}$\\
        Compute $l(\mathbf{d}|\mathbf{Z}^{t})$ (equation \eqref{eq:logL})\\
        \While {$\textit{t} < N_{draw}$}
        {
             Propose a new sample $\textbf{Z}_{\boldsymbol{\Phi}}\sp{\prime}$, $\textbf{Z}_{\boldsymbol{\eta}}\sp{\prime}$ and $\nu\sp{\prime}$ from proposal distribution $q(\textbf{Z}\sp{\prime},\textbf{Z}^{t-1})$\; 
             $\mathbf{\tilde X}_{\boldsymbol{\Phi}}^{\sp{\prime}}$, $\mathbf{\tilde X}_{\boldsymbol{\eta}_{app}}^{\sp{\prime}} \gets$ $\text{G}_{\boldsymbol{\Phi}}(\textbf{Z}_{\boldsymbol{\Phi}}\sp{\prime})$,  $\text{G}_{\boldsymbol{\eta}}(\textbf{Z}_{\boldsymbol{\eta}}\sp{\prime})$ \\
            Perform post-processing (section \ref{sec:training}):             $\boldsymbol{\Phi}^{\sp{\prime}}$, $\boldsymbol{\eta}_{app}^{\sp{\prime}}\gets \mathbf{\tilde X}_{\boldsymbol{\Phi}}^{\sp{\prime}}$, $\mathbf{\tilde X}_{\boldsymbol{\eta}_{app}}^{\sp{\prime}}$ and
            convert $\boldsymbol{\Phi}^{\sp{\prime}}$ into slowness $\textbf{s}^{\sp{\prime}}$ (equations \eqref{eq:keff}-\eqref{eq:slow})\\
             $\mathbf{d}_{sim} = g^{LF}(\mathbf{s}\sp{\prime}) + \nu\sp{\prime}  \boldsymbol{\eta}_{app}\sp{\prime}$\\
             Compute $l(\mathbf{d}|\mathbf{Z}\sp{\prime})$\\
             Compute probability of acceptance $\alpha \gets e^{(l(\textbf{d}|\textbf{Z}\sp{\prime}) - l(\textbf{d}|\textbf{Z}^{t-1}))}$\\
             Draw $U$ from a uniform distribution $\mathcal{U}(0,1)$\\
            \uIf {$U < \alpha$}
            {
                 $\textbf{Z}^{t} \gets \textbf{Z}\sp{\prime}$
            }
            \Else
            {
             $\textbf{Z}^{t} \gets  \textbf{Z}^{t-1}$
            }
            $\textit{t} = \textit{t}+1$\;
        }
        \Begin($\text{G}_{\boldsymbol{\Phi}} {(} \textbf{Z}_{\boldsymbol{\Phi}} {)}$){
             Performs a series of transposed convolution layers with pre-trained weights\;
             \Return {$\mathbf{\tilde X}_{\boldsymbol{\Phi}}$}
        }
        \Begin($\text{G}_{\boldsymbol{\eta}} {(} \textbf{Z}_{\boldsymbol{\eta}} {)}$){
             Performs a series of transposed convolution layers with pre-trained weights\;
             \Return {$\mathbf{\tilde X}_{\boldsymbol{\eta}_{app}}$}
        }
\end{algorithm}  
\vspace{0.2cm}


\section{Results}\label{sec:results}

In our numerical experiments we consider two parallel vertically-oriented boreholes, one containing $30$ sources and the other $30$ receivers.
The model domain on which the numerical experiment is performed is $4 \times 6.1$ m ($40 \times 61$ pixels). Sources and receivers are distributed evenly between $0.2$ and $6$ m depth in intervals of $0.2$ m and the two boreholes are located at $0$ m and $4$ m along the horizontal axis, respectively. In the straight-ray and eikonal forward solvers, the model domain is discretized evenly into $0.1$ m square cells. The FDTD simulated responses are performed using a spatial discretization of $0.025$ m and a time discretization of $0.15$ ns. The FDTD simulation requires the dielectric constant of the medium $\boldsymbol{\kappa}_{b}$ and electrical conductivity fields as input. We assume a constant conductivity of $0.002$ S/m across the model domain. The dielectric constant $\boldsymbol{\kappa}_{b}$ is obtained using equation \eqref{eq:keff}. The model-error databases corresponding to $\boldsymbol{\eta}^{eikonal-SR}$ and $\boldsymbol{\eta}^{FDTD-SR}$ contain $10,000$ images, each of which requires a  simulation using the low- and high-fidelity forward solvers. In the next subsections, we present results obtained from SGAN training and subsequent inversions.


\subsection{Quality assessment of generative models}\label{sec:resultsEik}

By training the SGAN on the subsurface model parameters and model error, we are able to reduce the two parameter spaces containing $2440$ and $900$ parameters (respectively) into two latent spaces, $\textbf{Z}_{\boldsymbol{\Phi}}$ and $\textbf{Z}_{\boldsymbol{\eta}}$, each of size $5 \times 5 \times 1$. In order to assess the quality of the generative models at a given training iteration, we calculate pixel-wise means and variances on a set of generated and training realizations. Based on this analysis, we found that the quality of the generated realizations could be further improved by scaling each realization by a spatially-varying correction factor intended to match the pixel-wise means of the TI's:

\begin{equation}\label{eq:mean_cor}
    \tilde{\textbf{X}} = G(\textbf{Z}) \cdot (\textbf{M}_{x} \oslash \textbf{M}_{\tilde{x}}),
\end{equation} 

\noindent
where $\textbf{M}_{x}$ is the mean of $10,000$ TI's, $\textbf{M}_{\tilde{x}}$ is the mean of $10,000$ SGAN realizations and $G(\textbf{Z})$ is a single SGAN realization to be corrected. The correction matrix obtained by element-wise division $\textbf{M}_{x} \oslash \textbf{M}_{\tilde{x}}$ contains the mean of generated SGAN realizations, and, thus, it is specific to a given training iteration. For the subsurface model-parameter realizations, we also evaluate the spatial auto-correlation within each realization by calculating directional semivariograms using the \textit{GSTools} package \citep{muler2020geostat}. 

\begin{figure}
    \centering
    \begin{minipage}{0.4\textwidth}
        \includegraphics[trim = 0 0 0 0, clip,width=\textwidth]{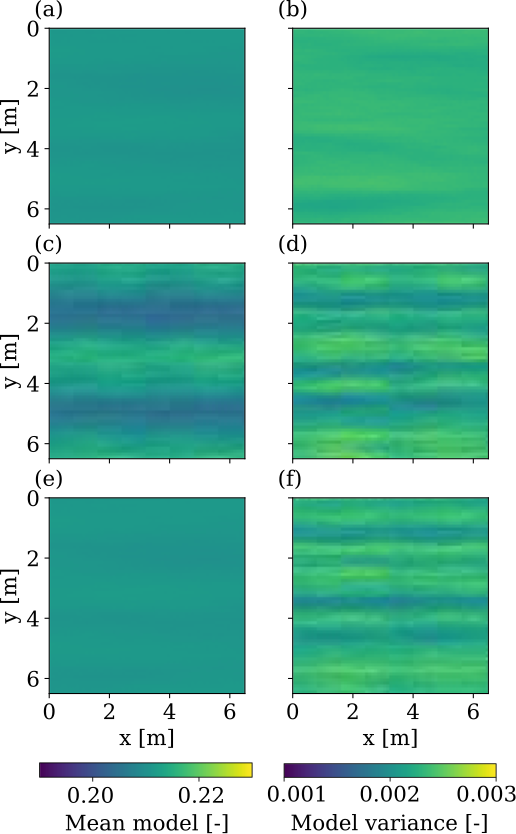}
    \end{minipage}
    \hfill
    \begin{minipage}{0.56\textwidth}
        \vspace{-7mm}
        \includegraphics[trim = 0 40 0 0, clip,width=\textwidth]{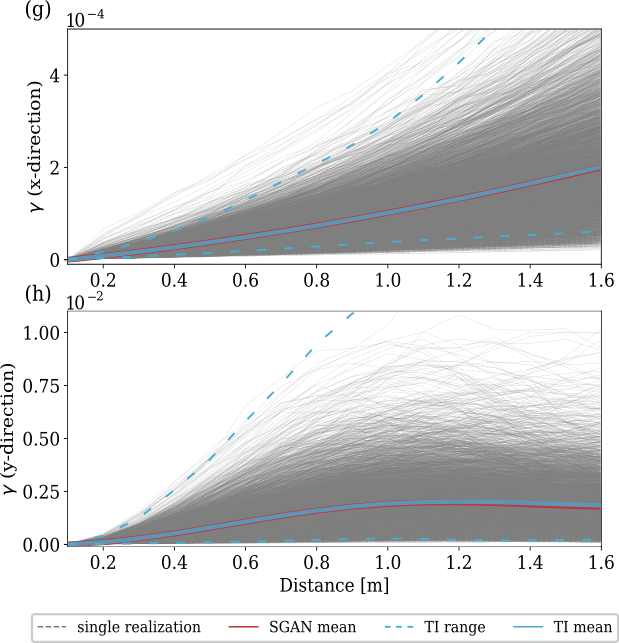}
        \includegraphics[trim = 0 0 0 0, clip,width=\textwidth]{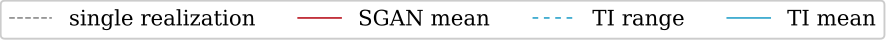}
    \end{minipage}
    \vspace{0.2cm}
    \caption{Statistics of subsurface model-parameter realizations after $58,000$ training iterations. Mean and variance images calculated on $5,000$: (a-b) TI realizations , (c-d) SGAN realizations before mean correction and (e-f) SGAN realizations after mean correction. The directional semivariograms in the (g) $x$- and (h) $y$-directions were calculated on $5,000$ TI realizations and SGAN realizations after mean correction. The gray lines are single semivariograms calculated on SGAN realizations after correction; their mean is marked as a solid red line and it is almost completely overlapped by the blue solid line, representing the mean of TI realizations. The blue dashed lines mark the TI realizations' range.}\label{fig:model}
\end{figure}

Training the SGAN for $58,000$ iterations with a batch containing $64$ images took about $8-9$ hours on one GPU GeForce GTX Titan X with $12$ GB memory. Figure \ref{fig:model} provides a comparison between the statistics of the subsurface model-parameter training images and SGAN realizations. We show the pixel-wise mean and variance of the TI's (Figs \ref{fig:model}a-b) and the SGAN realizations before (Figs \ref{fig:model}c-d) and after (Figs \ref{fig:model}e-f) applying the correction in equation \ref{eq:mean_cor}. The SGAN mean image before correction shows horizontal band-like features. After mean correction, this effect decreases and the image becomes closer to homogeneous. The variance images, however, do not exhibit the same improvement following the correction and look overall similar in both cases (Figs \ref{fig:model}d and f). The spatial statistics represented by the directional semivariograms in $x$- and $y$-directions are given in Figures \ref{fig:model}g and h, respectively; the mean semivariograms are calculated over $5,000$ TI (blue) and corrected SGAN (red) model realizations. The two mean curves fall on top of each other, indicating a good agreement between the TI and corrected SGAN realizations. Furthermore, the semivariograms of single SGAN realizations (gray curves) are mostly concentrated within the ranges of the TI (dashed blue curves).

\begin{figure}
    \centering
    \includegraphics[trim = 0 0 0 0,  clip,width=0.75\columnwidth]{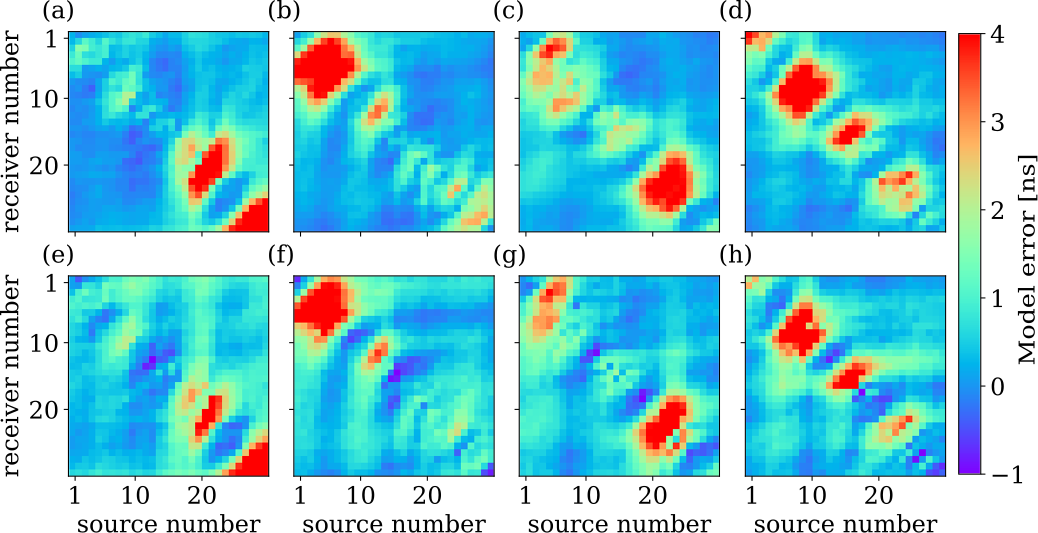}
    \caption{Examples of actual model-error realizations (a-d) $\boldsymbol{\eta}^{eikonal - SR}$ and (e-h) $\boldsymbol{\eta}^{FDTD - SR}$. Figures in the same column were calculated for the same subsurface model-parameter realization.} \label{fig:Err_sample}
\end{figure}

\begin{figure}
    \centering
    \includegraphics[trim = 0 0 0 0,  clip,width=0.8\columnwidth]{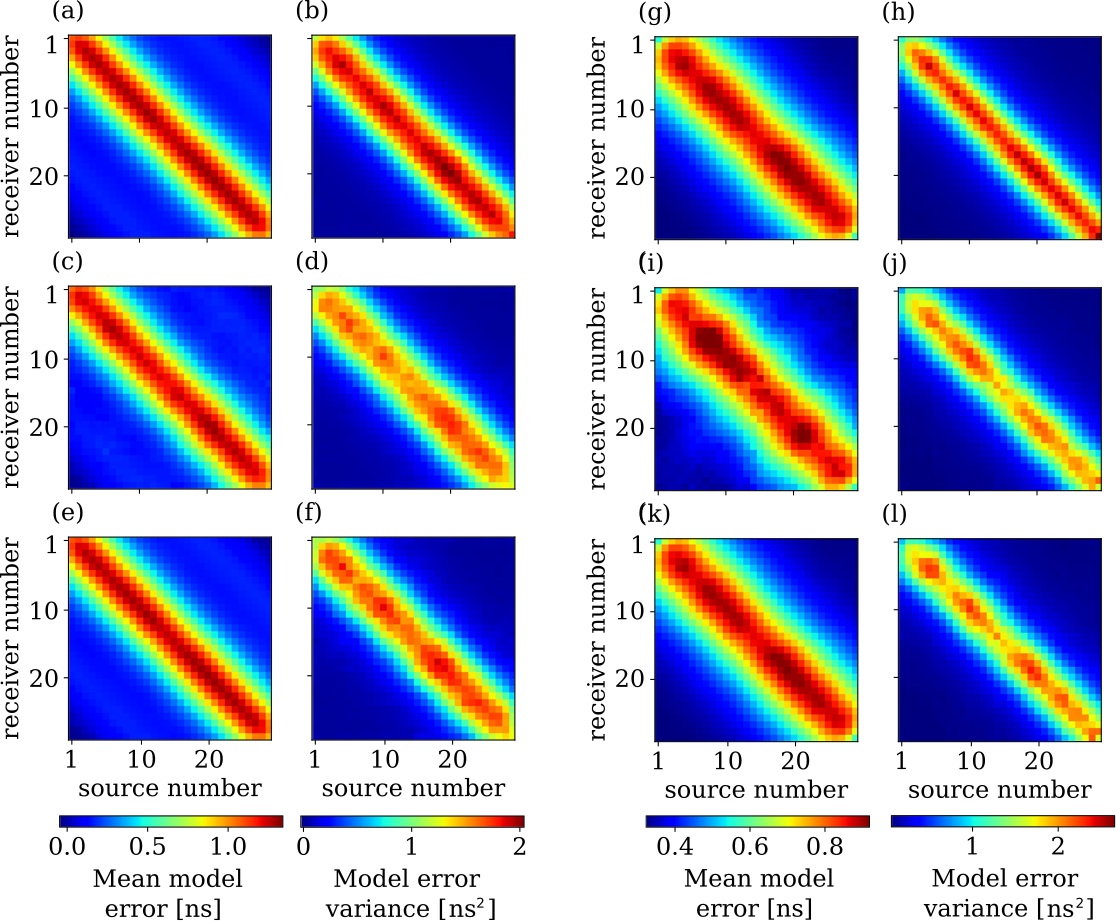}
    \caption{Model errors of (a-f) $\boldsymbol{\eta}^{eikonal - SR}$ and (g-l) $\boldsymbol{\eta}^{FDTD - SR}$. Pixel-wise mean and variance of $10,000$ (a-b and g-h) TI realizations, (c-d and i-j) SGAN realizations before mean correction and (e-f and k-l) SGAN realizations after mean correction (see equation \eqref{eq:mean_cor}).} \label{fig:Eik_err}
\end{figure}

A similar mean correction and assessment to those described above for the subsurface model-parameter SGAN training are performed for the model error training. Example model-error TI's for the two types of model errors considered in this paper are shown in Figure \ref{fig:Err_sample}. In most cases, $\boldsymbol{\eta}^{FDTD - SR}$ has a larger range of error values compared to $\boldsymbol{\eta}^{eikonal-SR}$ and displays similar features to $\boldsymbol{\eta}^{eikonal-SR}$ with additional off-diagonal patterns. Figure \ref{fig:Eik_err} provides a comparison between the pixel-wise mean and variance of the model-error TI's and those of the SGAN realizations before and after the mean correction. Although the mean image of the SGAN generated $\boldsymbol{\eta}^{eikonal-SR}_{app}$ realizations before correction (Fig. \ref{fig:Eik_err}c) is close to that of the TI realizations (Fig. \ref{fig:Eik_err}a), it underestimates the model-error mean on the diagonal. After correction (Fig. \ref{fig:Eik_err}e), the bias in the mean is removed and the variance (Fig. \ref{fig:Eik_err}f), which also suffers from underestimation on the diagonal, is slightly improved. Training with $\boldsymbol{\eta}^{FDTD - SR}$ realizations proved to be more challenging and required larger number of training iterations ($450,000$ iterations as opposed to $250,000$ for $\boldsymbol{\eta}^{eikonal - SR}$). The SGAN mean image before correction (Fig. \ref{fig:Eik_err}i) is distorted compared to the TI mean image (Fig. \ref{fig:Eik_err}g). We attribute this difference to the patchy nature of the $\boldsymbol{\eta}^{FDTD - SR}$ realizations and features that extend to elements further off-diagonal (Fig. \ref{fig:Err_sample}). These distortions were reduced after applying the mean correction (Fig. \ref{fig:Eik_err}k), although improvements in the variance (Fig. \ref{fig:Eik_err}l) are not as visible. One can observe a broken pattern on the diagonal in the $\boldsymbol{\eta}^{FDTD - SR}$ TI's mean and variance images (Figs. \ref{fig:Eik_err}g and h). This pattern can also be found in $\boldsymbol{\eta}^{eikonal - SR}$ TI's mean and variance images (Figs. \ref{fig:Eik_err}a and b), albeit to a lesser extent. Since the subsurface model-parameter TIs on which model errors are calculated were randomly chosen, we attribute this pattern to be a result of the forward modeling process rather than a repetitive pattern in the subsurface model-parameter TIs.

\begin{figure}
    \centering
        \includegraphics[trim = 0 0 0 0,  clip,width=0.5\textwidth]{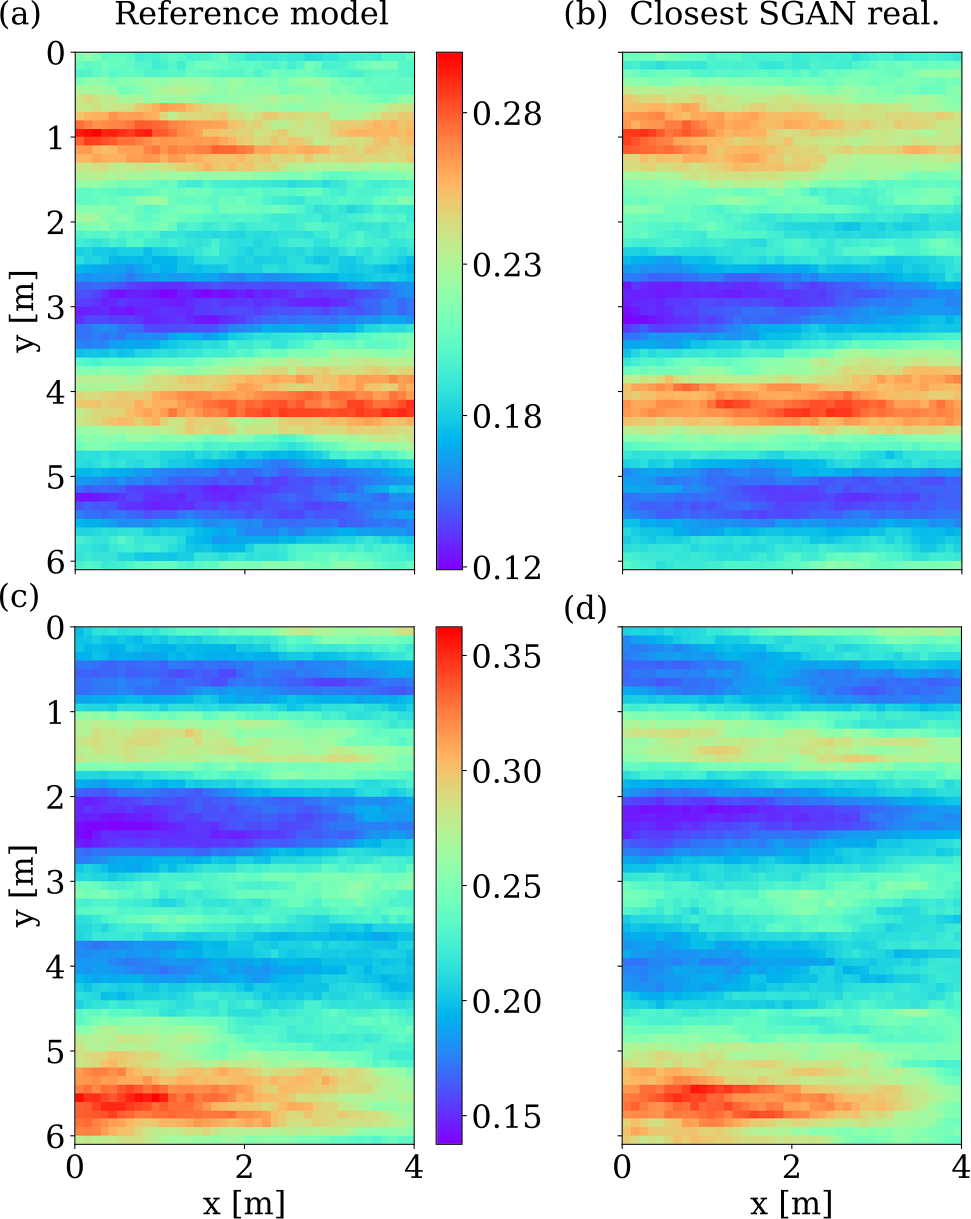}%
        \caption{Reference models (a) $1$ and (c) $2$ and (b and d) corresponding closest SGAN realizations obtained from pixel-to-pixel inversion considering 25 latent parameters.}\label{fig:closest_model}
\end{figure}

Finally, we test the ability of the SGAN to capture the true model by performing a pixel-to-pixel MCMC inversion (i.e., the actual pixel values are considered as data in the inversion) on two reference models, cropped out of the testing segment of the subsurface model-parameter TI described in Section \ref{sec:model}. We consider the maximum a posteriori estimate of pixel-to-pixel based inversion results as being the closest possible SGAN representation of the reference model ('closest SGAN realization'). Figure \ref{fig:closest_model} shows the considered reference models and their corresponding closest SGAN realization illustrating the capability of the SGAN to generate model realizations that closely resemble their reference models. 

\subsection{Inversion results}

We perform inversion of data generated from the two multi-Gaussian reference models in Figures \ref{fig:closest_model}a and \ref{fig:closest_model}c that we refer to as 'Model $1$' and 'Model $2$', respectively. The synthetic data for each reference model are created using the high-fidelity forward solver, which is either $g^{eikonal}$ or $g^{FDTD}$ depending on the type of model error considered (i.e., $\boldsymbol{\eta}^{eikonal - SR}$ or $\boldsymbol{\eta}^{FDTD - SR}$). The data are contaminated with random noise drawn from a normal distribution $\mathcal{N}(0, 0.5^{2}~ns^{2})$. We consider in our analysis only those traveltime data corresponding to source-receiver angles of less than $50^{\circ}$ from the horizontal, as is commonly done with field data to avoid borehole and antenna effects \citep{irving2005effect}. This leads to a total of $858$ traveltimes to be considered in the inversion. Note that the number (25) of subsurface model parameters $\textbf{Z}_{\boldsymbol{\Phi}}$ to be estimated is the same for all considered approaches. The SGAN-ME approach requires estimation of $26$ additional parameters: $25$ for the model error $\textbf{Z}_{\boldsymbol{\eta}}$ along with auxiliary parameter $\nu$. Note that the number of parameters in $\textbf{Z}_{\boldsymbol{\Phi}}$ and $\textbf{Z}_{\boldsymbol{\eta}}$ is chosen based on a trade-off between inversion performance and efficiency. It is chosen such that it remains low while enduring high-quality subsurface model estimation.   

For each of the considered approaches, we show the maximum a posteriori estimate. Given the uniform prior on the parameters, this corresponds also to the maximum-likelihood solution. For comparison, we calculate the root mean-square-error (RMSE) and structural similarity (SSIM) index for each approach including that of the closest SGAN realization obtained by pixel-by-pixel inversion. We consider two different RMSE values: one on the subsurface model parameters denoted by RMSE$_{\boldsymbol{\Phi}}$ and the other on the data denoted RMSE$_{\textbf{d}}$. The RMSE metric gives an indication as to the spread of residuals, with larger weight given to higher values, while the SSIM complements the latter by measuring the similarity of two images (here these are images of either the subsurface model parameters or model errors) in terms of their structure (see Appendix \ref{ap:measure}). The above metrics are calculated for the maximum-likelihood realization in the case of pixel-to-pixel inversion whereas in data-based inversions, they represent an average value for the last $50\%$ samples of the chains. In the case of the inferred model error, we also calculate what we refer to as "error recovery". This measure serves as an indication of how well the model error is approximated, by taking the average posterior mean-squared-error (MSE) between the approximated model error and the reference model error ($MSE(\boldsymbol{\eta}_{app},\boldsymbol{\eta}_{ref})$) and dividing it by the MSE between the reference model and $0$ ($MSE(\boldsymbol{\eta}_{ref},0)$). 

\subsubsection{Convergence}\label{sec:convergence}

We use the Gelman-Rubin diagnostic \citep{gelman1992inference} and declare convergence when all inferred parameters satisfy $\Hat{R} \leq 1.2$. The initial jump rate scaling factor was set to $5$ for all inversion runs. The minimum jump rate scaling factor had to be adjusted in each inversion individually in order to achieve a reasonable acceptance rate (ideally $20-30\%$ and not more than $50\%$) and convergence. A value of $0.2$ was often suitable to achieve convergence and reasonable acceptance rates with some SGAN-ME cases requiring slightly smaller values ($0.15-0.2$). In Table \ref{tab:converence}, we provide convergence information for each inversion approach. All inversions reached convergence, but the number of steps required differ between approaches. More steps are needed to reach convergence with the SGAN-ME approach. The mean acceptance rates in Table \ref{tab:converence} are consistently higher for the covariance approach compared to the other approaches due to its inflated error term, which increases the chance for proposed samples to be accepted in the MCMC. 

\begin{table}
 \caption{Inversion convergence for Test Case 1 ($\boldsymbol{\eta}^{eikonal-SR}$) and Test Case 2 ($\boldsymbol{\eta}^{FDTD-SR}$). The mean acceptance rate represents the average acceptance rate of the two tested reference models excluding the first $20,000$ steps.}
 \centering
 \begin{tabular}{P{2cm}P{2cm}P{2cm}P{2cm}P{2cm}}
 \hline
 \multirow{2}{*}[0.5em]{\specialcell{Model\\ error \\ type}} &  \multirow{2}{*}{\specialcell{Inversion \\ approach}} & \multicolumn{2}{c}{\specialcell{Nr. of \\MCMC steps \\ (per chain)}} & \multirow{2}{*}[1.0em]{\specialcell{Mean \\acceptance \\ rate (excl. \\ burn-in) $[\%]$}}\\ [0.5ex] 
   \cline{3-4} 
   & & {Model $1$} & {Model $2$}  & \\ [0.5ex] 
   \hline   
   \multirow{4}{*}{$\boldsymbol{\eta}^{eikonal-SR}$} &straight-ray & $95,510$ & $22,060$ & $28$\\ 

   & Covariance & $43,860$ & $189,760$ & $36$ \\ 

   &SGAN-ME & $108,960$ & $382,620$ & $18$ \\

   &eikonal& $34,710$ & $61,160$ & $23$ \\ 
   
   \hline
   \multirow{3}{*}{$\boldsymbol{\eta}^{FDTD-SR}$} &straight-ray & $53,160$ & $141,110$ & $18$ \\ 

   &Covariance & $27,810$ & $188,010$  & $35$ \\ 

   &SGAN-ME & $363,760$ & $437,810$ & $23$ \\

   \hline

 \end{tabular}
 \label{tab:converence}
\end{table}


\subsubsection{Test Case 1: eikonal - straight-ray model error}\label{subsec:Eikinversion}

\begin{figure}
    \centering
        \includegraphics[trim = 0 0 0 0,  clip,width=\textwidth]{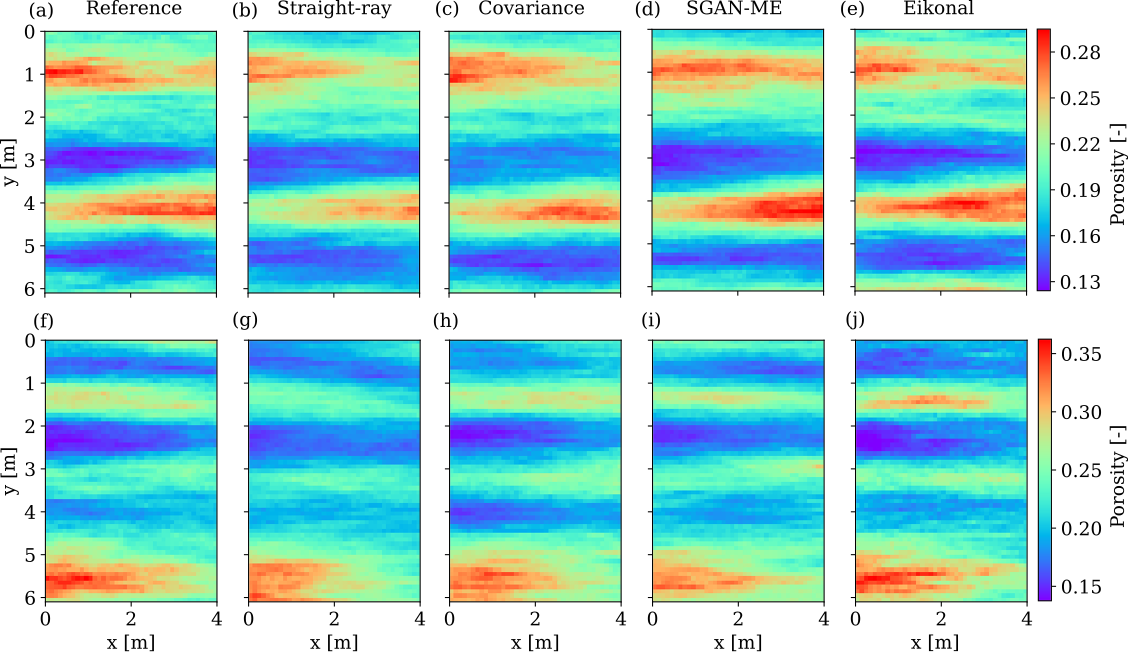}%
        \caption{Inversion results for reference Models (a) $1$ and (f) $2$ for Test Case 1  ($\boldsymbol{\eta}^{eikonal-SR}$). (b)-(e) and (g)-(j) are the  maximum-likelihood realizations obtained from inversion using the straight-ray, covariance, SGAN-ME and eikonal approaches. The first three approaches use the straight-ray solver for the forward response during inversion, while the observed data for all approaches were created using the eikonal solver.}\label{fig:post50Eik}
\end{figure}

\begin{figure}
    \centering
    \includegraphics[trim = 0 0 0 0,  clip,width=\textwidth]{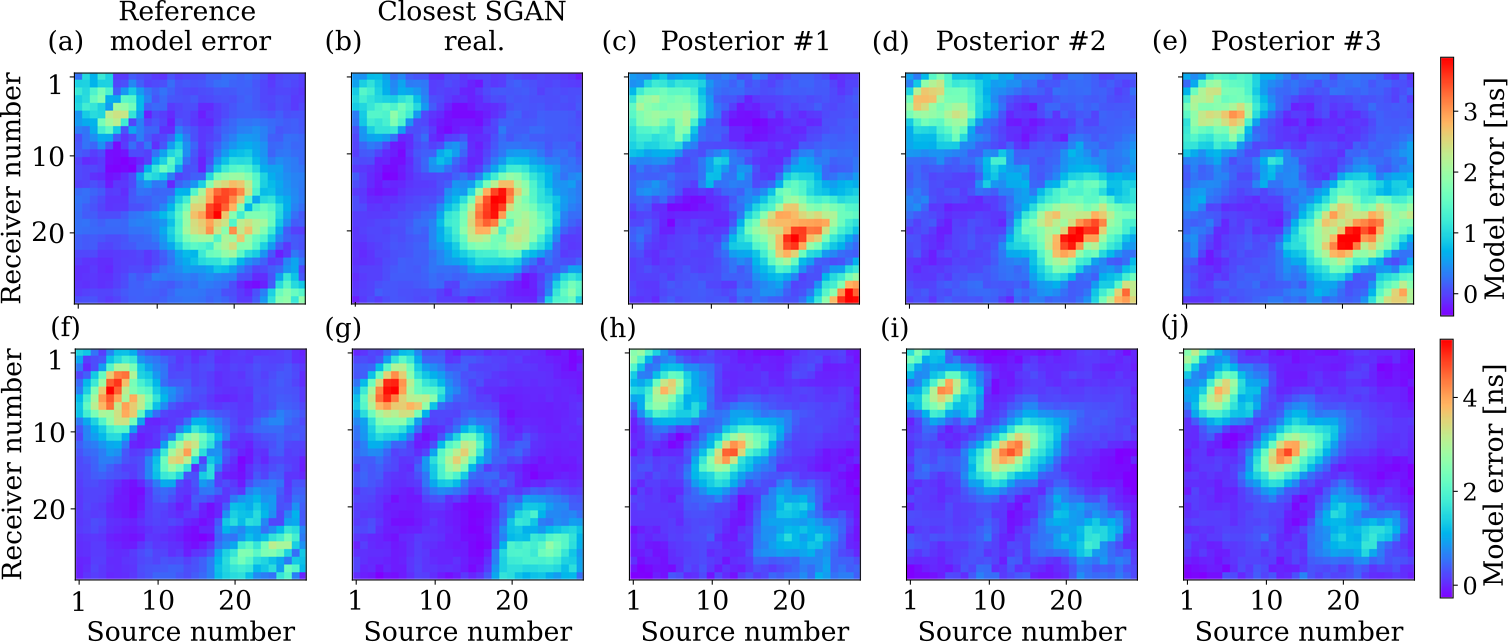}%
    \caption{Model errors for Test Case 1 ($\boldsymbol{\eta}^{eikonal-SR}$) representing the discrepancy between the eikonal and straight-ray solvers. (a) and (f) are reference model errors calculated based on reference Models $1$ and $2$ in Figures \ref{fig:post50Eik}a and \ref{fig:post50Eik}f, respectively, (b) and (g) are the corresponding closest SGAN model-error realizations obtained from pixel-to-pixel inversion and (c)-(e) and (h)-(j) are posterior samples obtained from inversion with the SGAN-ME approach.}\label{fig:error50Eik}
\end{figure}

We first consider inversion results with model error $\boldsymbol{\eta}^{eikonal - SR}$ in terms of maximum-likelihood solutions of the straight-ray, covariance, SGAN-ME and eikonal-based inversion approaches in Figure \ref{fig:post50Eik} and RMSE$_{\boldsymbol{\Phi}}$, SSIM and RMSE$_{\textbf{d}}$ in Table \ref{tab:eikonal}. Generally speaking and given values in Table \ref{tab:eikonal}, the SGAN-ME approach exhibits better overall performance compared to the straight-ray and covariance approaches, scoring lower RMSE and higher SSIM values. The SGAN-ME approach captures well the general structure of the various porosity zones in both test models. The spatial representation of model errors in Figure \ref{fig:error50Eik} together with values in Table \ref{tab:error}, suggest that SGAN-ME is able to recover a large part of the model error (about $51\%$ for Model $1$ and $67\%$ for Model $2$). Table \ref{tab:error} also indicates that the closest SGAN realizations obtained by the pixel-based inversions consistently reached better scores than the closest of the $10,000$ model realizations used for training, thereby indicating that the SGAN generalizes well for the model error.

We now consider results for Model $1$ specifically. The SGAN-ME and eikonal solutions exhibit similar structures between $0$ and $5$ m depth, resulting in similar SSIM values ($0.79$ and $0.78$, respectively). The low porosity zone between $5$ and $5.5$ m depth is thinner in the SGAN-ME solution and the high-porosity zone between $4-4.5$ is overestimated. This can be explained by considering the SGAN-ME model-error posterior samples in Figures \ref{fig:error50Eik}c-e. Although the features on the diagonal (and close to diagonal) are correctly located, they are underestimated for source-receiver pairs $(15,15)-(20,20)$ and overestimated for source-receiver pairs $(20,20)-(25,25)$ causing overestimation of porosity in the region corresponding to the latter source-receiver pairs. Furthermore, the model errors at the bottom right corner of all posterior samples in Figure \ref{fig:error50Eik}c-e are overestimated and differ by up to $\sim 2$ ns from the truth, translating to a thicker high-porosity layer at the bottom of the subsurface model ($5.5$-$6$ m). The covariance solution overestimates the low-porosity zone at around $3$ m depth. It scores the same RMSE$_{\boldsymbol{\Phi}}$ as the straight-ray solution ($0.016$) but receives higher SSIM ($0.74$ versus $0.72$) and slightly lower RMSE$_{\textbf{d}}$ ($0.75$ ns versus $0.77$ ns) scores.

As for Model $2$, the SGAN-ME maximum-likelihood realization is the only solution properly reconstructing the porosity structure between $0$-$1$ m depths. Other approaches, including the eikonal solution do not have a clear layered structure around these depths. The eikonal solution tend to overestimate some high-porosity zones ($4$-$4.5$ m in Model $1$ and around $1.5$-$1.7$ m in Model $2$) and exhibit rough texture in its solution to Model $2$. The covariance solution underestimates the porosity at $4$ m depth but still surpasses the straight-ray solution in both subsurface model-parameters scores (RMSE$_{\boldsymbol{\Phi}}$ and SSIM). As opposed to Model $1$, here the straight-ray solution fits the data significantly better than the covariance solution (RMSE$_{\textbf{d}}$ of $0.87$ ns for straight-ray versus $1.17$ ns for covariance). Furthermore, the straight-ray solutions are smooth and do not contain major artifacts. They do however, generally underestimate high-porosity zones and receive the highest RMSE and lowest SSIM scores in most cases. 

\begin{figure}
    \centering
    \includegraphics[trim = 0 20 0 20,  clip,width=0.47\textwidth]{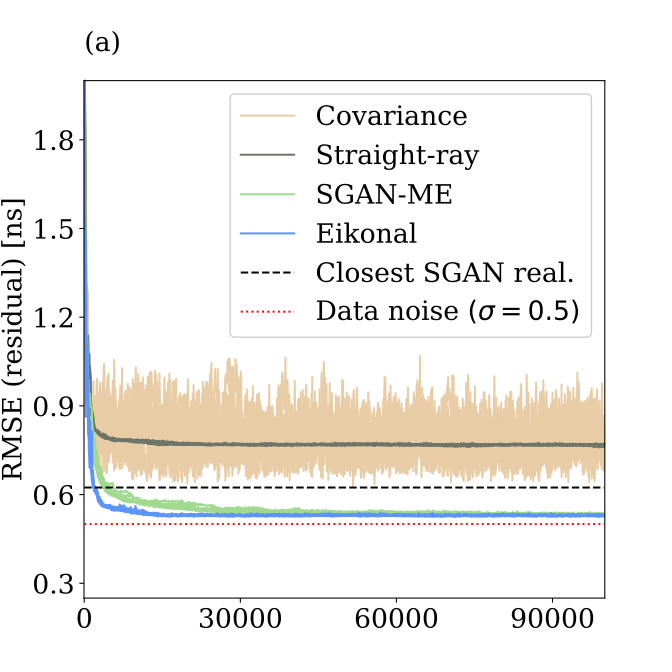}%
    \includegraphics[trim = 0 20 0 20,  clip,width=0.47\textwidth]{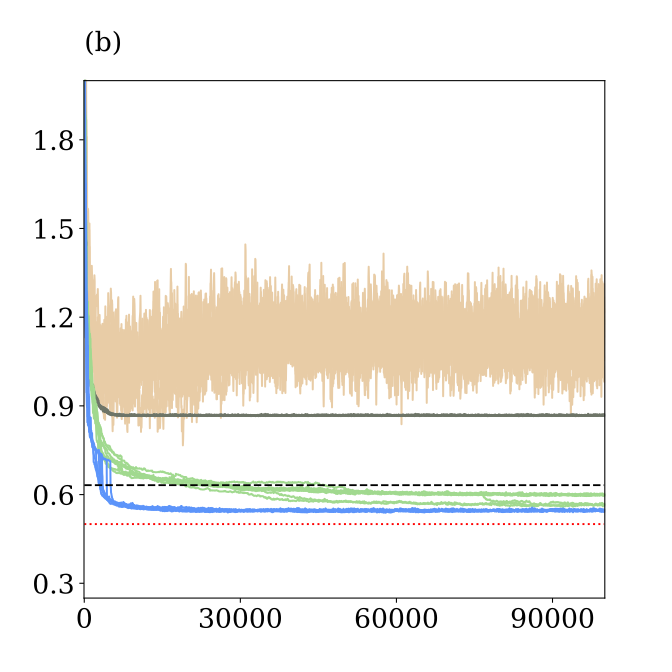}\\%
    \includegraphics[trim = 0 5 0 20,  clip,width=0.47\textwidth]{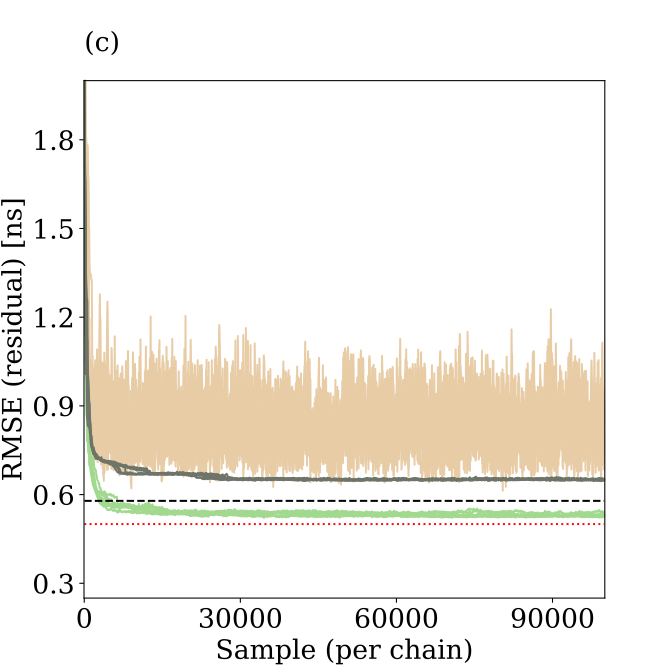}%
    \includegraphics[trim = 0 5 0 20,  clip,width=0.47\textwidth]{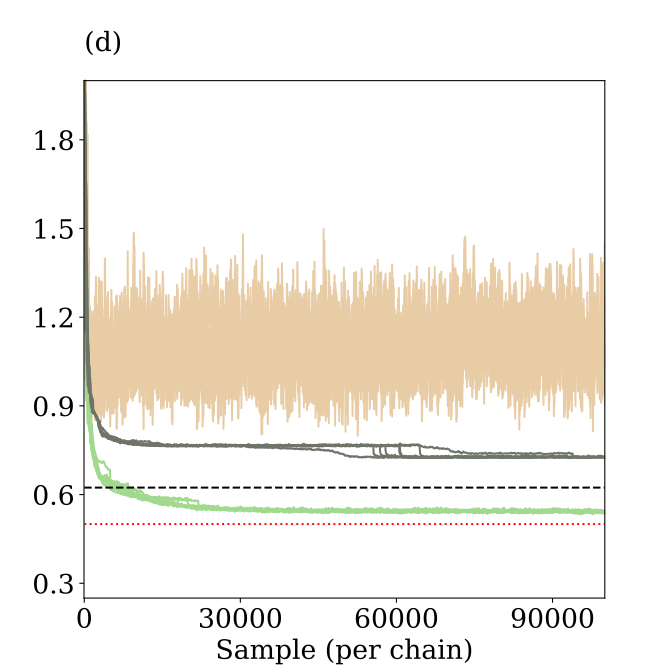}
    \caption{Data fit (RMSE$_{\textbf{d}}$) for inversion considering: modelling error $\boldsymbol{\eta}^{eikonal - SR}$ for reference models (a) $1$ and (b) $2$ and modelling error $\boldsymbol{\eta}^{FDTD - SR}$ for reference models (c) $1$ and (d) $2$.}\label{fig:data_fit}
\end{figure}

As can be seen in Table \ref{tab:eikonal}, the RMSE$_{\textbf{d}}$ was also calculated for the closest SGAN realization using the high-fidelity forward solver, namely the eikonal solver. For better visualization, we show in Figure \ref{fig:data_fit} the RMSE$_{\textbf{d}}$ values of each approach and for each of its eight chains along $100,000$ sequential samples (per chain). The data fit plots corresponding to Model $1$ and $2$ and model error $\boldsymbol{\eta}^{eikonal-SR}$ (Figs \ref{fig:data_fit}a and b) indicate that our SGAN-ME approach fits the data as well as the eikonal solver, close to the noise level of $0.5$ ns (indicated by the red dotted line) and significantly better than the straight-ray and covariance approaches. The RMSE$_{\textbf{d}}$ of the closest SGAN realization (indicated by a dotted black line) is higher compared to that of the eikonal and SGAN-ME inversion approaches, but lower than that of the straight-ray and covariance approaches.

\begin{figure}
    \centering
    \includegraphics[trim = 0 0 0 0,  clip,width=\textwidth]{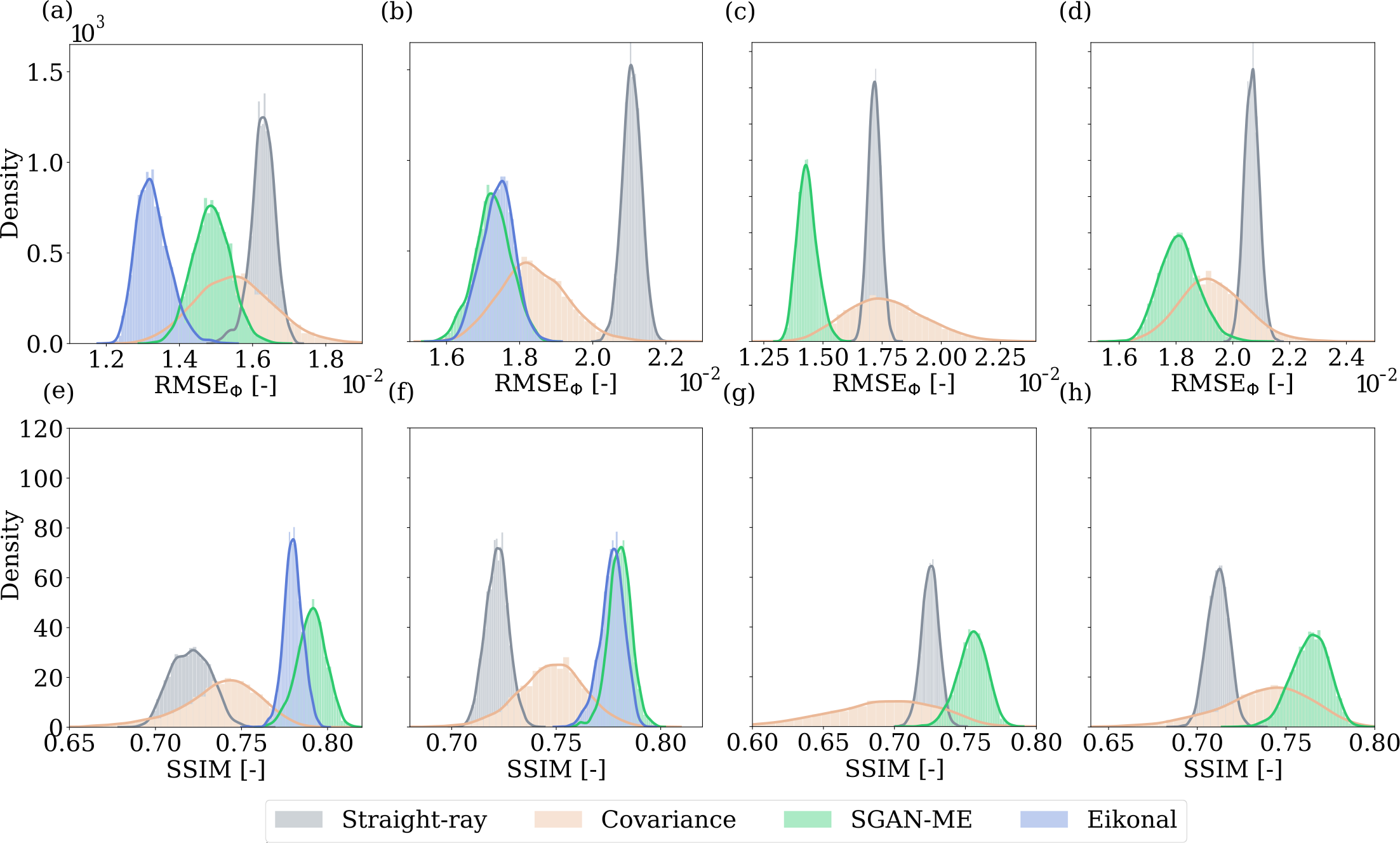}%
    \caption{RMSE$_{\boldsymbol{\Phi}}$ and SSIM distributions of posterior samples for inversion considering: Test Case 1 ($\boldsymbol{\eta}^{eikonal - SR}$) for reference models (a and e) $1$ and (b and f) $2$ and Test Case 2 ($\boldsymbol{\eta}^{FDTD - SR}$) for reference Models (c and g) $1$ and (d and h) $2$. The high-fidelity solution is only available in Test Case 1 (blue area in a, b, e and f).}\label{fig:SSIM_mse}
\end{figure}

Finally, we represent posterior samples in the form of RMSE$_{\boldsymbol{\Phi}}$ and SSIM distributions (Figs. \ref{fig:SSIM_mse}a,b,e,f). The RMSE$_{\boldsymbol{\Phi}}$ and SSIM values, calculated separately for each posterior sample, were plotted as a normalized density function to which a Gaussian kernel was fitted. It is observed that the SGAN-ME approach generally results in RMSE$_{\boldsymbol{\Phi}}$ and SSIM distributions that rank higher than the straight-ray and covariance approaches. For Model 2, the RMSE$_{\boldsymbol{\Phi}}$ and SSIM distributions associated with SGAN-ME almost completely overlap those corresponding to the model error-free eikonal approach. The SGAN-ME posterior distributions are characterized by intermediate widths as opposed to the covariance approach for which RMSE$_{\boldsymbol{\Phi}}$ and SSIM values vary widely and to the straight-ray approach for which the distribution is narrow and with the worst statistics. 

\begin{table}
\centering
 \caption{Inversion results for Test Case 1  ($\boldsymbol{\eta}^{eikonal-SR}$) in terms of the subsurface model considering $g^{LF} = g^{SR}$ and $g^{HF} = g^{eikonal}$. The RMSE$_{\boldsymbol{\Phi}}$ and SSIM values are average values of the posterior samples. The RMSE$_{\boldsymbol{\Phi}}$ of each posterior sample was calculated on porosity values with respect to the corresponding reference model. The SSIM was calculated on normalized images in the range of $\left[0, 1\right]$. The SSIM can take values between -1 and 1, where 1 indicates identical images. The RMSE$_{\textbf{d}}$ represents the data fit with respect to the observed data and is an average value over the last draws from the eight MCMC chains. For more details see Appendix \ref{ap:measure}.}
 
 \begin{tabular}{c c c c c} 
 \hline\hline
  Model &Inv. approach & RMSE$_{\boldsymbol{\Phi}}$ [-] & SSIM [-] & RMSE$_{\textbf{d}}$ [ns]\\ [0.5ex] 
   \hline
   &True & $0$ & $1$ & $0.5$\\
   \hline
    
  \multirow{5}{*}{1} &straight-ray & $0.016$ & $0.72$ & $0.77$ \\ 

   & Covariance & $0.016$ & $0.74$ &  $0.75$ \\ 

   &SGAN-ME & $0.015$ & $0.79$ & $0.53$ \\

   &eikonal& $0.013$ & $0.78$ & $0.53$ \\ 
   
   &Closest SGAN real.& $0.010$ & $0.83$ & $0.62$ \\ 
   
  \hline


  \multirow{5}{*}{2}&straight-ray& $0.021$ & $0.72$ & $0.87$ \\ 

  &Covariance & $0.018$ & $0.75$ & $1.17$\\ 

  &SGAN-ME& $0.017$ & $0.78$ & $0.55$ \\

  &eikonal& $0.017$ & $0.78$ & $0.55$\\ 
  
  &Closest SGAN real.& $0.013$ & $0.83$ & $0.63$ \\ 
  \hline\hline
 \end{tabular}
 \label{tab:eikonal}
\end{table}

\begin{table}
\centering
 \caption{Inversion results in terms of model-error estimation for the two considered reference models ($1$ and $2$) and Test Case 1 ($\boldsymbol{\eta}^{eikonal-SR}$) and 2 ($\boldsymbol{\eta}^{FDTD-SR}$) . The given RMSE and SSIM values are average values of the posterior samples of model errors. The RMSE of each posterior sample was calculated with respect to the corresponding reference model error. The SSIM was calculated on normalized images in the range of $\left[0, 1\right]$. The SSIM can take values between -1 and 1, where 1 indicate identical images. The error recovery represents the fraction of mean-squared-error (MSE) of posterior samples MSE$(\boldsymbol{\eta}_{app},\boldsymbol{\eta}_{ref})$ compared to the MSE$(\boldsymbol{\eta}_{ref},0)$ of the reference model with respect to $0$ and can range between $0\%$ to $100\%$. For more details see Appendix \ref{ap:measure}.}
 \begin{tabular}{c c c c c c} 
 \hline\hline
  Model error & Model &Inv. approach & RMSE [ns] & SSIM [-] & Error recovery [\%]\\ [0.5ex] 
   \hline
    & &True & $0$ & $1$ & $100$\\
    \hline

   \multirow{6}{*}{$\boldsymbol{\eta}^{eikonal-SR}$}&\multirow{2}{*}{1}&SGAN-ME & $0.67$ & $0.56$ & $51$ \\ 
   &&Closest SGAN real.& $0.23$ & $0.87$ & $94$ \\
   &&Closest database real.& $0.29$ & $0.87$ & $90$ \\
   \cline{2-6}

   &\multirow{2}{*}{2}& SGAN-ME & $0.66$ & $0.64$ & $67$ \\
   &&Closest SGAN real.& $0.29$ & $0.86$ & $94$ \\
   &&Closest database real.& $0.54$ & $0.63$ & $77$ \\
  \hline\hline
     
   \multirow{6}{*}{$\boldsymbol{\eta}^{FDTD-SR}$}&\multirow{2}{*}{1}&SGAN-ME & $0.49$ & $0.68$ &$74$ \\ 
   & &Closest SGAN real.& $0.24$ & $0.88$ &$94$ \\ 
   &&Closest database real.& $0.33$ & $0.85$ & $88$ \\
   
   \cline{2-6}

   &\multirow{2}{*}{2}&SGAN-ME & $0.63$ & $0.72$ & $71$ \\ 
   &&Closest SGAN real.& $0.32$ & $0.89$ &$92$\\
   &&Closest database real.& $0.56$ & $0.71$ & $78$ \\
  \hline\hline
 \end{tabular}
 \label{tab:error}
\end{table}

\subsubsection{Test Case 2: FDTD - straight-ray model error}\label{subsec:inversion}

\begin{figure}
    \centering
    \includegraphics[trim = 0 0 0 0,  clip,width=0.82\textwidth]{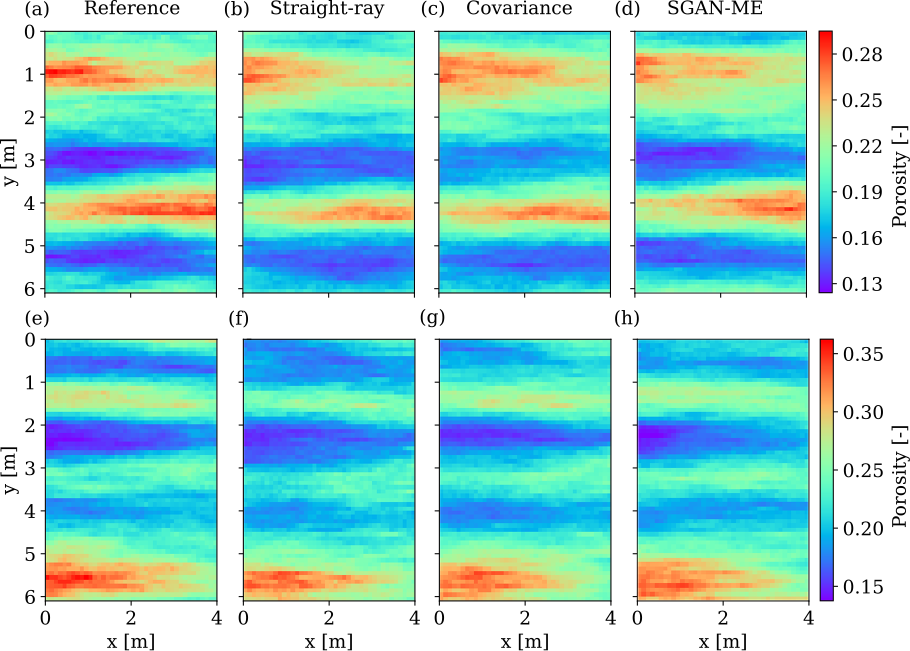}%
    \caption{Inversion results for reference models (a) $1$ and (e) $2$ for Test Case 2 ($\boldsymbol{\eta}^{FDTD-SR}$). (b)-(d) and (f)-(h) are the maximum-likelihood realizations obtained from inversion using the straight-ray, covariance and SGAN-ME approaches. All three approaches use the straight-ray solver for the forward response during inversion, while the observed data were created using the FDTD solver.}\label{fig:post50FW}
\end{figure}

We now consider the model error $\boldsymbol{\eta}^{FDTD - SR}$ for the same reference porosity models and create the synthetic data using the FDTD forward solver. Here we compare only between the straight-ray, covariance and SGAN-ME approaches due to the excessive computational time needed to perform MCMC inversion with the FDTD forward solver \citep{hunziker2019}. Results for this test case can be found in Figure \ref{fig:post50FW} and Table \ref{tab:FWSR} which show the maximum-likelihood solution of the straight-ray, covariance and SGAN-ME approaches for the two reference models and their respective RMSE and SSIM scores.

The maximum-likelihood solution together with the RMSE$_{\boldsymbol{\Phi}}$ and SSIM values in Table \ref{tab:FWSR} suggest that the SGAN-ME results capture both the magnitude and structure of porosity for Model $1$ and is the closest to values observed for the closest SGAN realization, having RMSE$_{\boldsymbol{\Phi}}$ of $0.0014$ ns (versus $0.010$ ns) and SSIM value of $0.75$ (versus $0.83$). The SGAN-ME approach is also able to recover large portions of the model error ($74\%$ error recovery) for this reference model (Table \ref{tab:error}). The high-porosity zone between $0.5$ and $1.5$ m depth is wider in the all solutions compared to reference Model $1$, although less visible in the SGAN-ME solution. Similarly, as was found previously for the case of $\boldsymbol{\eta}^{eikonal - SR}$, the covariance solution consistently overestimates the porosity around $3$ m depth for Model $1$ (Figs. \ref{fig:post50Eik}c and \ref{fig:post50FW}c). All compared approaches underestimate the high porosity zone between $\sim3.8$-$4.5$ m depth and overestimate the low-porosity zone between $5$-$5.5$ m depth.

\begin{figure}
    \centering
    \includegraphics[trim = 0 0 0 0,  clip,width=\textwidth]{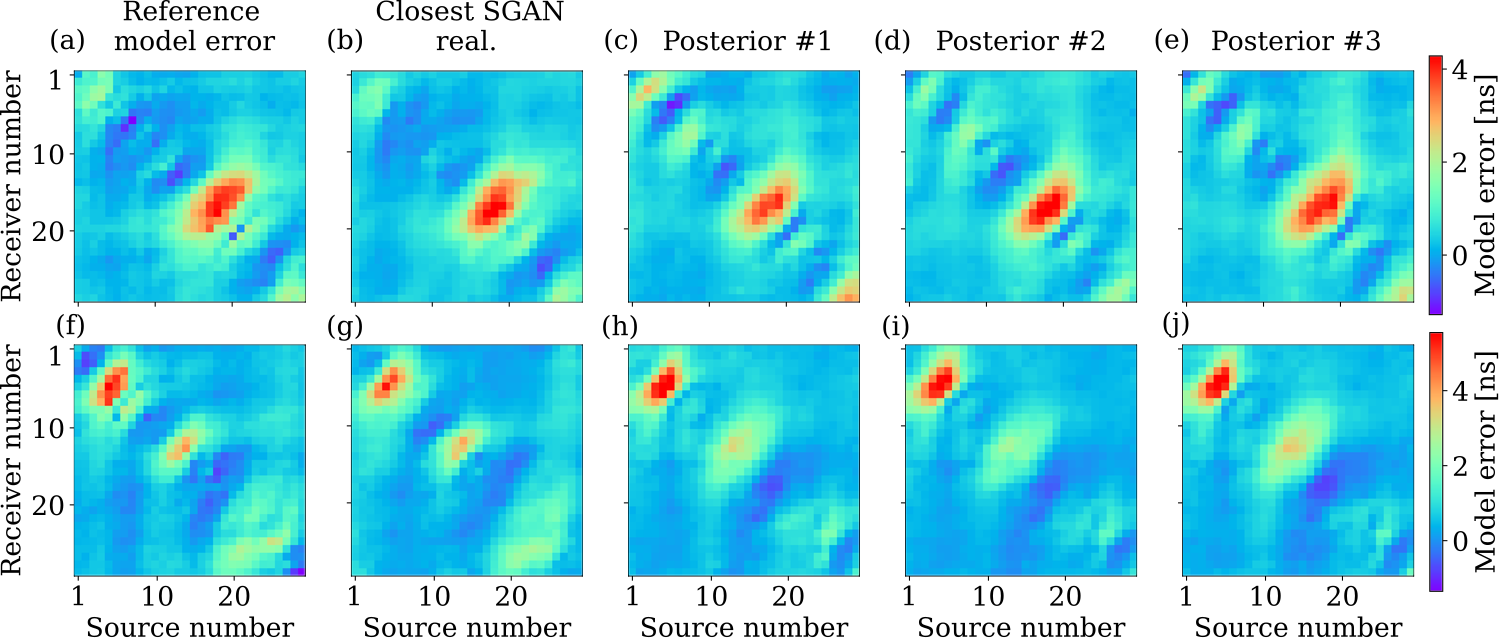}
    \caption{Model errors for Test Case 2 ($\boldsymbol{\eta}^{FDTD-SR}$) representing the discrepancy between the FDTD and straight-ray solvers. (a) and (f) are reference model errors calculated based on reference models $1$ and $2$ in Figures \ref{fig:post50FW}a and \ref{fig:post50FW}e, respectively, (b) and (g) are the corresponding closest SGAN model error realizations obtained from pixel-to-pixel inversion and (c)-(e) and (h)-(j) are three posterior samples obtained from inversion with the SGAN-ME approach.}\label{fig:error50FW}
\end{figure}

As for Model $2$, the porosity structure between $0$ and $1$ m is better defined in the SGAN-ME solution compared to the other approaches. The porosity zone between $1.8$ and $2.8$ m depth is overestimated in the right hand side of the SGAN solution. This part of the subsurface model is covered by receivers $10$-$15$. Indeed, the posterior samples displayed in Figure \ref{fig:error50FW}h-j show a larger diagonal feature between receivers $10$-$15$ and sources $10$-$15$ than in the reference model error for those source-receiver pairs. Nonetheless, the inferred SGAN-ME model error recovers $71\%$ of the true model error (Table \ref{tab:error}). 

For both reference models, the RMSE$_{\boldsymbol{\Phi}}$ of the covariance and straight-ray approaches are increasing or remain the same when going from $\boldsymbol{\eta}^{eikonal - SR}$ to $\boldsymbol{\eta}^{FDTD - SR}$. Interestingly, the SGAN-ME inversion result corresponding to Model $1$ improves from $\boldsymbol{\eta}^{eikonal - SR}$ to $\boldsymbol{\eta}^{FDTD - SR}$, with RMSE$_{\boldsymbol{\Phi}}$ decreasing from $0.015$ to $0.014$ while the SSIM value decreases from $0.79$ to $0.75$. This improvement in RMSE$_{\boldsymbol{\Phi}}$ score can be linked to better error recovery, which increases from $51\%$ for $\boldsymbol{\eta}^{eikonal - SR}$ to $74\%$ for $\boldsymbol{\eta}^{FDTD - SR}$. Notice that in both types of model errors the closest SGAN model-error realizations obtained by pixel-based inversion (Figs. \ref{fig:error50Eik}b and \ref{fig:error50Eik}g for $\boldsymbol{\eta}^{eikonal - SR}$ and Figs. \ref{fig:error50FW}b and \ref{fig:error50FW}g for $\boldsymbol{\eta}^{FDTD - SR}$) strongly resemble their reference model errors and their error recovery is between $92$ to $94\%$, further exemplifying the ability of the SGAN to represent model errors. Tables \ref{tab:eikonal} and \ref{tab:FWSR} and Figure \ref{fig:data_fit} show that the SGAN-ME approach is able to fit the data equally well in both test cases and approaches the noise contamination level. Finally, we observe that the posterior samples in the form of RMSE$_{\boldsymbol{\Phi}}$ and SSIM distribution (Figure \ref{fig:SSIM_mse}c,d,g,h) show similar patterns as for Test Case 1, in the sense that the SGAN-ME approach generally results in RMSE$_{\boldsymbol{\Phi}}$ and SSIM distributions that rank higher than the straight-ray and covariance approaches. Again, the SGAN-ME distributions are characterized with intermediate widths as opposed to the covariance approach for which RMSE$_{\boldsymbol{\Phi}}$ and SSIM values vary widely and to the straight-ray approach for which the distribution is narrow and exhibits the worst statistics. 

\begin{table}
\centering
 \caption{Inversion results for Test Case 2  ($\boldsymbol{\eta}^{FDTD-SR}$) in terms of the subsurface model considering $g^{LF} = g^{SR}$ and $g^{HF} = g^{FDTD}$. The RMSE$_{\boldsymbol{\Phi}}$ and SSIM values are average values of the posterior samples. The RMSE$_{\boldsymbol{\Phi}}$ of each posterior sample was calculated on porosity values with respect to the corresponding reference model. The SSIM was calculated on normalized images in the range of $\left[0, 1\right]$. The SSIM can take values between -1 and 1, where 1 indicates identical images. The RMSE$_{\textbf{d}}$ represents the data fit with respect to the observed data and is an average value over the last draws from the eight MCMC chains. For more details see appendix \ref{ap:measure}.}
 
 \begin{tabular}{c c c c c} 
 \hline\hline
  Model &Inv. approach & RMSE$_{\boldsymbol{\Phi}}$ [-] & SSIM [-] & RMSE$_{\textbf{d}}$ [ns] \\
   \hline
   &True & $0$ & $1$ & $0.5$\\ 
     \hline
    
  \multirow{4}{*}{1}&straight-ray & $0.017$ & $0.73$ & $0.65$ \\ 

  &Covariance & $0.018$ & $0.69$ & $0.84$ \\ 

   &SGAN-ME & $0.014$ & $0.75$ &$0.53$ \\
   
   &Closest SGAN real.& $0.010$ & $0.83$ & $0.58$ \\
  \hline


  \multirow{4}{*}{2}&straight-ray& $0.021$ & $0.71$ & $0.73$ \\

  &Covariance & $0.019$ & $0.74$ & $1.16$ \\ 

  &SGAN-ME& $0.018$ & $0.76$ & $0.54$ \\
  &Closest SGAN real.& $0.013$ & $0.83$ & $0.62$ \\ 
  
  \hline\hline
 \end{tabular}
 \label{tab:FWSR}
\end{table}


\section{Discussion}

Our results demonstrate the suitability of our SGAN architecture and training procedure to represent model errors and the ability of SGAN-ME inversions to infer them for a given subsurface model realization (Figs. \ref{fig:error50Eik} and \ref{fig:error50FW}). Among the considered inversion methods employing a low-fidelity forward solver, the SGAN-ME inversion scored RMSE (${\boldsymbol{\Phi}}$ and ${d}$) and SSIM values that are the closest to those obtained when the high-fidelity forward eikonal solver is used in the inversion (Table \ref{tab:eikonal}). This indicates that inferring the model error during inversion using the SGAN-ME offers an overall better performance compared to ignoring model errors or accounting for them by inflating the error term in the likelihood function following \citet{hansen2014accounting}. Somewhat surprisingly, the straight-ray approach, where model errors are neglected, resulted in subsurface models with relatively minor artifacts (Figs. \ref{fig:post50Eik}b, \ref{fig:post50Eik}g, \ref{fig:post50FW}b and \ref{fig:post50FW}f). This is likely a consequence of the SGAN dimensionality reduction. The dimensionality of the subsurface model domain is reduced in our examples from $2440$ parameters to $25$ latent parameters, thus, limiting strong artifacts at the expense of the ability to achieve high likelihoods. We expect that more artifacts would appear when inverting the data in the original high-dimensional subsurface model space. 

In all tested cases, the SGAN-ME is able to infer meaningful model-error representations (Figs. \ref{fig:error50Eik} and \ref{fig:error50FW}) ranging between $71$ and $74\%$ recovery of the true model error in the $\boldsymbol{\eta}^{FDTD-SR}$ Test case (Table \ref{tab:error}). By jointly inferring the subsurface model parameters and the model error, SGAN-ME enables identification and localization of regions in the subsurface model that are prone to large model errors. Some of the inferred model errors are still misplaced (Figure \ref{fig:error50Eik}c-e) or underestimated (Figure \ref{fig:error50Eik}h-j). This could suggest that the inferred model error accommodates inadequacies between the subsurface-model realizations that can be generated by the SGAN and the reference subsurface model used to generate the data. Indeed, with 25 parameters it is of course impossible to fully represent all the geostatistical variability of our training image. Tables \ref{tab:eikonal} and \ref{tab:FWSR} reinforce this hypothesis, as they show that the closest SGAN realization obtained from a pixel-to-pixel inversion does not fit the data as well as the eikonal or our SGAN-ME approach, implying a certain bias in the SGAN-ME inversions. A possible solution to address this problem would be to perform a hierarchical inversion in which the standard deviation of the data error is one of the inferred parameters \citep{Malinverno2004Hierarchical}. Initial results with such an hierarchical approach have been inconclusive to date and require further investigation. 

The RMSE$_{\textbf{d}}$ values corresponding to SGAN-ME are very similar to those obtained when using the high-fidelity eikonal solver (Table \ref{tab:eikonal}). For both types of errors, $\boldsymbol{\eta}^{eikonal-SR}$ and $\boldsymbol{\eta}^{FDTD-SR}$, SGAN-ME is found to fit the data significantly better than the straight-ray and covariance approaches with values close to the noise level of $\sigma = 0.5$ ns (Tables \ref{tab:eikonal} and \ref{tab:FWSR}). We have seen that the impact of the type of model-error size on data fit is small in the SGAN-ME approach, indicating its robustness in fitting the data by inferring the model error. The covariance approach is characterized by a large variability of RMSE$_{\textbf{d}}$ values throughout the inversion due to the inflation of the likelihood function, and hence a wide range of realizations are accepted. This variability in model realizations is also observed in Figure \ref{fig:SSIM_mse}, where the covariance-based RMSE$_{\boldsymbol{\Phi}}$ and SSIM distributions exhibit the largest variance. The straight-ray approach spans a smaller range of posterior realizations, but those present poor RMSE$_{\boldsymbol{\Phi}}$ and SSIM scores. In that regard, the SGAN-ME presents a combination of small uncertainty (intermediate posterior widths) and the best RMSE$_{\boldsymbol{\Phi}}$ and SSIM scores.

In agreement with other approaches treating model errors as the discrepancy between a low- and a high-fidelity solver, we stress that our method is unable to quantify any model errors arising from simplifications in the high-fidelity solver or an inappropriate prior model (training data) of the subsurface properties. As a deep learning method, our approach depends on the availability of training data (i.e. subsurface-model representation and two fidelity-varying forward solvers). Note also that the networks are model and model-error specific, meaning that new training is required if considering a different set-up. Furthermore, our SGAN-ME approach combines multiple nonlinear transformations leading to MCMC convergence issues. Here, we relied on the $DREAM_{(ZS)}$ algorithm and found that convergence was highly sensitive to the chosen jump-rate scaling factor. In the future, it would be beneficial to assess if convergence could be improved by using other MCMC samplers such as gradient-, Hamiltonian-dynamics- \citep{duane1987hybrid,neal2011mcmc} or diffusion- \citep{roberts1996exponential,roberts1998optimal} based samplers.


\section{Conclusions}
We present a methodology accounting for model errors in Bayesian inversion using deep generative neural networks. In contrast to most existing methods, our approach makes no restrictive Gaussian assumptions about the statistical distribution of the model errors arising from using a fast low-fidelity solver instead of a slow high-fidelity solver. We use SGANs to learn two separate generative models: one for the subsurface model parameters of interest and the other for the model errors. The underlying low-dimensional latent parameterizations are then used to jointly infer the subsurface model parameters and model error via MCMC using the fast low-fidelity forward solver, thereby, allowing for significant speed-up. By doing so, we are able to improve the posterior estimates of subsurface model parameters and model errors. Our SGAN-ME method is shown to perform better than in cases where model errors are ignored or accounted for using a Gaussian error model. In fact, the quality of the posterior solutions is close to results obtained when using a high-fidelity forward solver in the MCMC. By providing posterior distributions of the model errors, it is possible to visualize where model errors occur and to identify regions where inversion results might be less reliable. This information could be used to locally replace low-fidelity simulations with high-fidelity simulations. Our focus has been on model errors due to simplified physics, but our approach and the extension discussed above could also be useful when considering coarse meshes for the forward computations. In addition, our approach could be extended to other fields of geophysics, for example, full-waveform inversion. Even if our SGAN-ME method works well in the considered test examples, we highlight the need to address MCMC instabilities due to the underlying nonlinearity of the SGAN transformation. Since the performance of our approach depends on the quality of the SGAN realizations, there is a need to further advance network architectures and training procedures for both subsurface model parameters and model errors. Further improvements could also be made by training the subsurface model and model error jointly with shared latent parameters or by combining our SGAN-ME approach with deep-learning based surrogate modeling.


\section{Acknowledgements}

This research was supported by the Swiss National Science Foundation (project number: \href{http://p3.snf.ch/project-184574}{184574}). We thank associated editor Juan-Carlos Afonso, as well as reviewers Brent Wheelock and Jianwei Ma for their constructive comments. The SGAN and MCMC scripts as well as test examples can be found in the following GitHub repository: \url{https://github.com/ShiLevy/SGAN_ME}.

\section*{Citation}
This is a pre-copyedited, author-produced PDF of an article accepted for publication in \textit{Geophysical Journal International} following peer review. The version of record\\

\noindent
Shiran Levy, Jürg Hunziker, Eric Laloy, James Irving, Niklas Linde, Using deep generative neural networks to account for model errors in Markov chain Monte Carlo inversion, \textit{Geophysical Journal International}, 2021;, ggab391, https://doi.org/10.1093/gji/ggab391\\

\noindent
is available online at: \url{https://academic.oup.com/gji/advance-article/doi/10.1093/gji/ggab391/6374556}

\bibliographystyle{apalike}
\bibliography{template}  

\newpage

\appendix

\section{Details on SGAN Architecture and training}\label{Ap:training}

Below we discuss the SGAN architecture and provide practical information about its training. 

\subsection{Network architecture}\label{ap:lr}
\begin{figure}[H]
    \centering
    \includegraphics[trim = 0 0 0 0,  clip,width=\textwidth]{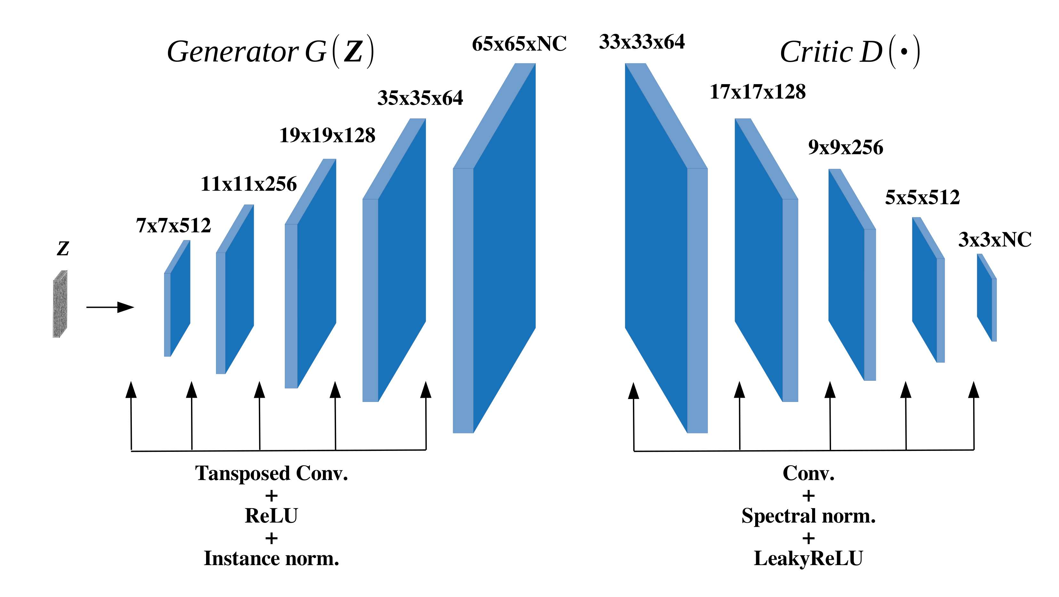}%
    \caption{SGAN architecture showing the activation and normalization types and output size after each convolution (/transposed convolution) with NC being the number of image channels (e.g. three channels in RGB images). When training over model errors the critic layers include mean-spectral-normalization as opposed to spectral normalization alone for subsurface-model training.}\label{fig:SGAN_arc_appendix}
\end{figure}

Figure \ref{fig:SGAN_arc_appendix} details the architecture of the SGAN used in this study. The learning rate of the generator (ratio of $1:4$ in learning rate between generator and critic) in subsurface-model training is $5e-05$ while it is $1e-06$ in model-error training. We found that using such a low learning rate was essential to avoid artifacts from appearing in the generated images. We used a batch size of $64$ even if a batch size of $32$ provides similar results. The hyper-parameters of each layer are detailed in Table \ref{tab:SGAN_param} and include the kernel, stride and padding sizes. We use the RMSProp \citep{tieleman2012lecture} optimizer in both generator and critic to update the parameters of the network.

\subsection{Effective receptive field and feature size}

\begin{table}
\centering
 \caption{SGAN hyper-parameters.}
 \begin{tabular}{c c c c c }
 \hline\hline
  & layer & kernel & stride & padding\\
   \hline
  \multirow{5}{*}{Generator}& $1$ & $5$ & $2$ & $3$\\ 

  &$2$ & $5$ & $2$ & $3$ \\ 

   &$3$ & $5$ & $2$ & $3$ \\
   
   &$4$ & $5$ & $2$ & $3$ \\
   
   &$5$ & $5$ & $2$ & $4$ \\
  \hline
  \multirow{5}{*}{Critic}& $1$ & $5$ & $2$ & $2$ \\ 
  
  &$2$ & $5$ & $2$ & $2$\\ 

   &$3$ & $5$ & $2$ & $2$ \\
   
   &$4$ & $5$ & $2$ & $2$ \\
   
   &$5$ & $1$ & $2$ & $0$ \\
  
  \hline\hline
 \end{tabular}
 \label{tab:SGAN_param}
\end{table}

A distinct difference between SGANs and GANs is the way information in the latent space is being translated into the image space. GANs usually involve a latent space vector where each latent parameter affects the resulting images globally, while in SGANs the latent parameters are ordered within a 2D/3D tensor and contain local information which overlaps in the image space. One of the limitations arising from using spatially-dependent information within a convolutional network is that a change in the dimensions of the latent space affects the output image size (see eq. \eqref{eq:output}). This means that the network output size is determined by the dimensions of the latent space. All input to the critic in SGANs must have the same dimensions, therefore, the dimensions of the TIs should match those of the generated images. 

\begin{figure}
    \centering
    \includegraphics[trim = 0 0 0 0,  clip,width=0.8\textwidth]{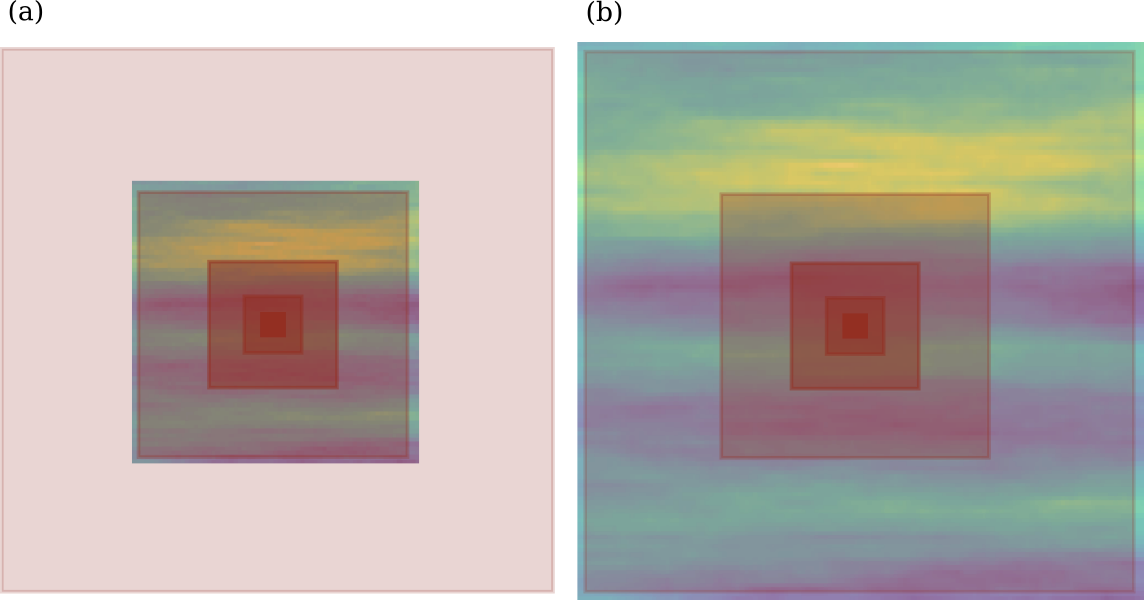}
    \caption{(a) Multi-Gaussian TI of dimensions $65x65$ pixels and (b) the same TI interpolated into $129x129$ pixels, both overlaid by the ERF's of neurons computed for $5$ sequential convolutional layers. The ERF is computed given $k=5$ and $s=2$ for all layers.}\label{fig:ERF}
\end{figure}

We can easily match image sizes by performing an interpolation on the TI to match the generated image dimensions (or vice versa). Note though that there is an indirect effect of image interpolation on the learning process that is related to the effective receptive field (ERF). The ERF is the area in the input (or output in the case of a generator) influencing a neuron in a given convolutional layer. The ERF is a function of the kernel and stride sizes and can be computed for the $l^{th}$ layer in the following way \citep{le2017receptive}:

\begin{equation}
    R_{l} = R_{l-1} + (k_{l}-1)\prod^{l-1}_{i=1} s_{i},
\end{equation}

where $R_{l}$ and $R_{l-1}$ are the ERF's of a neuron in the current and previous layers, $k_{l}$ is the kernel size in the current layer, $s_{i}$ is the stride in layer $i$ and $R_{0}=1$. Although the ERF size does not depend on the size of the image or latent space, an interpolation to the TI affect the network for given kernel and stride sizes. The reason is that for an interpolated TI, features within the image are larger/smaller and therefore, the portion of the features seen by a neuron is changed (see Figure \ref{fig:ERF}). As illustrated in Figure \ref{fig:ERF}, where the ERFs of $5$ layers are plotted on top of a TI before and after interpolation for a given network architecture, the resolution in which neurons in each layer 'see' features of difference scales changes with interpolation. This means that some scales cannot be properly resolved which can lead to a mode collapse or a failure of the network to learn the underlying data distribution.

Hence, it is important to test how well the output/input image is covered by the ERF's of neurons in different layers. Since the SGAN was proven to be substantially more sensitive to changes in $k$ or $s$ than in $p$ (padding; see section \ref{sec:training}), in our work we limited the generated image size using padding when we increased the number of latent parameters.  

\section{Quality measure calculation}\label{ap:measure}

Here we expand the information concerning the quantitative measures appearing in Tables \ref{tab:eikonal}, \ref{tab:error} and \ref{tab:FWSR}. We use RMSE as a metric for model and data fit. The RMSE of the model ($RMSE_{\boldsymbol{\Phi}}$) is calculated on porosity values of individual posterior realizations (only the last $50\%$ of each chain is considered) with respect to the reference model: 

\begin{equation}
RMSE_{\boldsymbol{\Phi}}=\sqrt{ \frac{\sum^{N_{\boldsymbol{\Phi}}}_{n=1} (\boldsymbol{\Phi}_{ref}-\boldsymbol{\Phi}_{n})^{2}}{N_{\boldsymbol{\Phi}}}},
\end{equation}

\noindent
where $N_{\boldsymbol{\Phi}}$ is the number of subsurface model parameters. The final reported $RMSE_{\boldsymbol{\Phi}}$ is the average value of posterior samples. The data RMSE ($RMSE_{d}$) is the average RMSE value in the last draw of the MCMC chains

\begin{equation}
RMSE_{\textbf{d}}=\sqrt{ \frac{\sum^{N_{d}}_{n=1}
(\textbf{d}-\textbf{d}^{sim}_{n})^{2}}{N_{d}}},
\end{equation}

\noindent
where $N_{d}$ is the number of data points. 

The structural similarity (SSIM; \citealp{wang2004image}) index of two images $U$ and $V$ is a common quantitative measure in image processing. It is calculated using sliding windows $\textbf{u}$ and $\textbf{v}$ of dimension $M \times M$ (we use a $7 \times 7$ window) subsampling the $[0,1]$ normalized images, 

\begin{equation}
SSIM(\textbf{u},\textbf{v}) = \frac{(2\mu_{\textbf{u}}\mu_{\textbf{v}} +C_{1})(2\sigma_{\textbf{uv}} +C_{2})}{(2\mu_{\textbf{u}}^{2} + \mu_{\textbf{v}}^{2} +C_{1})(2\sigma_{\textbf{u}}^{2} + \sigma_{\textbf{v}}^{2} +C_{2})},
\end{equation}

\noindent
where $\mu_{\textbf{u}}$ and $\mu_{\textbf{v}}$ are the mean values over $\textbf{u}$ and $\textbf{v}$, $\sigma_{\textbf{u}}^{2}$ and $\sigma_{\textbf{v}}^{2}$ are the respective variances of $\textbf{u}$ and $\textbf{v}$ and $\sigma_{\textbf{uv}}$ is the covariance between $\textbf{u}$ and $\textbf{v}$. We follow \citet{wang2004image} and set $C_{1}=0.01$ $C_{2}=0.03$.

The error recovery value is calculated based on MSE values of the reference model error with respect to $0$ ($MSE(\boldsymbol{\eta}_{ref},0)$) and the MSE of the inferred model error with respect to the reference model ($MSE(\boldsymbol{\eta},\boldsymbol{\eta}_{ref})$): 

\begin{equation}
MSE(\boldsymbol{\eta}_{ref},0) = \frac{\sum_{n=1}^{N_{\boldsymbol{\eta}}} (0 - \boldsymbol{\eta}_{ref,n})^{2}}{N_{\boldsymbol{\eta}}},
\end{equation}

\begin{equation}
MSE(\boldsymbol{\eta}_{app},\boldsymbol{\eta}_{ref}) = \frac{\sum_{n=1}^{N_{\boldsymbol{\eta}}} (\boldsymbol{\eta}_{ref}-\boldsymbol{\eta}_{app,n})^{2}}{N_{\boldsymbol{\eta}}},
\end{equation}

\noindent
where $N_{\boldsymbol{\eta}}$ is the number of model error parameters. The  error recovery is the fraction of the average $MSE(\boldsymbol{\eta}_{app},\boldsymbol{\eta}_{ref})$ within posterior samples and $MSE(\boldsymbol{\eta}_{ref},0)$ given in percentage:

\begin{equation}
ER = \frac{\overline{MSE}(\boldsymbol{\eta}_{app},\boldsymbol{\eta}_{ref})}{MSE(\boldsymbol{\eta}_{ref},0)} * 100\%.
\end{equation}

\end{document}